\RequirePackage[2020-02-02]{latexrelease}
\documentclass[conference]{IEEEtran}
\usepackage{algorithm,algpseudocode}
\usepackage{graphicx}
\usepackage{tabularx}
\usepackage{amsmath,amssymb,amsfonts}
\usepackage{hyperref}
\usepackage{booktabs}
\usepackage{makecell}
\usepackage{subcaption}
\usepackage{cite}
\usepackage[printwatermark]{xwatermark}
\usepackage{listings}
\usepackage[available,functional,reproduced]{ndssbadges}

\usepackage{cleveref}
\usepackage{xcolor,pifont,xspace}
\usepackage{tikz}
\usepackage{color, colortbl}
\newcommand{\ournameNoSpace}{NeuroStrike} 
\newcommand{\ourname}{\ournameNoSpace\xspace}

\begin{document}
\title{NeuroStrike: \\Neuron-Level Attacks on Aligned LLMs}
\pagestyle{plain} %
\setcounter{page}{1}
\makeatletter
\newcommand{\linebreakand}{%
  \end{@IEEEauthorhalign}
  \hfill\mbox{}\par
  \mbox{}\hfill\begin{@IEEEauthorhalign}
}
\makeatother

\author{\IEEEauthorblockN{Lichao Wu}
	\IEEEauthorblockA{Technical University of Darmstadt\\
		lichao.wu@trust.tu-darmstadt.de}
	\and
	\IEEEauthorblockN{Sasha Behrouzi}
	\IEEEauthorblockA{Technical University of Darmstadt\\
		sasha.behrouzi@trust.tu-darmstadt.de}  
	\and
	\IEEEauthorblockN{Mohamadreza Rostami}
	\IEEEauthorblockA{Technical University of Darmstadt\\
		mohamadreza.rostami@trust.tu-darmstadt.de}
  
	\linebreakand
	\IEEEauthorblockN{Maximilian Thang}
	\IEEEauthorblockA{Technical University of Darmstadt\\
		maximilian.thang@stud.tu-darmstadt.de} 
	\and
	\IEEEauthorblockN{Stjepan Picek}
	\IEEEauthorblockA{University of Zagreb \& Radboud University\\
		stjepan.picek@ru.nl} 
	\and
	\IEEEauthorblockN{Ahmad-Reza Sadeghi}
	\IEEEauthorblockA{Technical University of Darmstadt\\
		ahmad.sadeghi@trust.tu-darmstadt.de}
        }

\IEEEoverridecommandlockouts
\makeatletter\def\@IEEEpubidpullup{6.5\baselineskip}\makeatother
\IEEEpubid{\parbox{\columnwidth}{
		Network and Distributed System Security (NDSS) Symposium 2026\\
		23 - 27 February 2026, San Diego, CA, USA\\
		ISBN 979-8-9919276-8-0\\  
		https://dx.doi.org/10.14722/ndss.2026.230660\\
		www.ndss-symposium.org
}
\hspace{\columnsep}\makebox[\columnwidth]{}}

\maketitle

\begin{abstract}
Safety alignment is critical for the ethical deployment of large language models (LLMs), guiding them to avoid generating harmful or unethical content. Current alignment techniques, such as supervised fine-tuning and reinforcement learning from human feedback, remain fragile and can be bypassed by carefully crafted adversarial prompts. Unfortunately, such attacks rely on trial and error, lack generalizability across models, and are constrained by scalability and reliability.

This paper presents NeuroStrike, a novel and generalizable attack framework that exploits a fundamental vulnerability introduced by alignment techniques: the reliance on sparse, specialized safety neurons responsible for detecting and suppressing harmful inputs. 
We apply NeuroStrike to both white-box and black-box settings: 
In the \emph{white-box setting}, NeuroStrike identifies safety neurons through feedforward activation analysis and prunes them during inference to disable safety mechanisms. 
In the \emph{black-box setting}, we propose the first LLM profiling attack, which leverages safety neuron transferability by training adversarial prompt generators on open-weight surrogate models and then deploying them against black-box and proprietary targets.
We evaluate NeuroStrike on over 20 open-weight LLMs from major LLM developers. By removing less than 0.6\% of neurons in targeted layers, NeuroStrike achieves an average attack success rate (ASR) of 76.9\% using only vanilla malicious prompts. Moreover, Neurostrike generalizes to four multimodal LLMs with 100\% ASR on unsafe image inputs. Safety neurons transfer effectively across architectures, raising ASR to 78.5\% on 11 fine-tuned models and 77.7\% on five distilled models. The black-box LLM profiling attack achieves an average ASR of 63.7\% across five black-box models, including Google’s Gemini family.

\end{abstract}

\IEEEpeerreviewmaketitle

\section{Introduction}

Large Language Models (LLMs) have dramatically transformed natural language processing, exhibiting extraordinary capabilities in tasks ranging from language generation and translation to complex reasoning and interactive dialogues~\cite{thirunavukarasu2023large,liu2025fin,kaddour2023challenges}. Despite these advancements, their extensive deployment across various industries raises significant security and safety concerns, notably the potential for generating harmful, misleading, or unsafe content~\cite{wei2023jailbroken}. To address these issues, techniques referred to as \emph{safety alignment} have been introduced. Implemented through post-training fine-tuning, safety alignment methods like Reinforcement Learning from Human Feedback (RLHF)~\cite{ouyang2022training} fine-tune models to align outputs with human ethical judgments, compressing harmful responses.

However, recent research has revealed significant limitations in current safety alignment methods for LLMs.
First, alignment mechanisms lack robustness; even benign fine-tuning intended to enhance general performance can inadvertently weaken existing safety constraints~\cite{qi2023fine}. Second, despite efforts to guide models toward ethical outputs, they remain susceptible to adversarial prompts, known as jailbreaks, which bypass safety mechanisms and elicit harmful responses~\cite{wei2023jailbroken,niu2024jailbreaking,shen2024anything}. Yet, crafting universally effective jailbreak prompts remains challenging, as differences in training data, model architectures, and alignment strategies severely limit their transferability, rendering existing offensive research largely ad hoc and empirical.
On the other hand, recent studies have attempted to interpret the safety mechanisms in LLMs either at the layer level~\cite{li2024safety} or at the feature level~\cite{chen2024finding}. However, these methods may not accurately pinpoint the critical components responsible for safety behaviors as they implicate nearly 10\% of model parameters as safety-related. Defensive technique~\cite{zhao2024towards} narrowly focuses on specific layers and is validated for limited LLMs, constraining its practical applicability across diverse/multimodal LLMs. These gaps highlight the urgent need for a deeper, principled understanding of the underlying mechanisms governing safety alignment, which could inform more targeted, reliable, and generalizable attacks.

\noindent
\textbf{Safety Alignment as a Loophole:} 
When analyzing the behavior of aligned LLMs, we identify an analogy between safety alignment and adversarial attacks~\cite{xu2024ban,qiu2019review,li2022backdoor}, where models exhibit predictable yet abnormal responses upon receiving specially crafted inputs. The aligned models are conditioned to respond predictably (e.g., ``I’m sorry, I cannot assist with that.'') to malicious inputs, thereby implicitly creating a \emph{safety trigger}. Inspired by neural interpretability research, which demonstrates that sophisticated behaviors in neural networks often originate from sparse, highly specialized neuron groups~\cite{bau2017network,tamkin2023codebook}, we hypothesize that safety alignment is similarly implemented via dedicated neurons, denoted as \emph{safety neurons}. Similar to how the human brain has neurons that help us distinguish right from wrong, LLMs rely on specific safety neurons to recognize and suppress harmful behavior. These neurons act as internal detectors, discriminating malicious inputs from benign queries by producing distinctive activation patterns. If an adversary accurately identifies and manipulates these safety neurons, either by suppressing their activation with carefully crafted input or directly pruning them, the safety-aligned model can be neutralized. This neutralization enables the direct elicitation of harmful outputs, bypassing the model’s intended safety alignment mechanisms.

\noindent
\textbf{Our Goals and Contributions:}
We present \ourname, a novel attack framework that analyzes and exploits the safety triggers introduced by the safety alignment. \ourname exploits insights from safety neurons' behavior to compromise both open-weight\footnote{Open-weight LLMs offer publicly available pre-trained weights independent of data or code openness.} and black-box (including proprietary) LLMs. Our framework leverages lightweight neuron activation analysis to identify safety neurons during inference, then removes or bypasses them for the attack. Our approach achieves high success rates for eliciting harmful outputs and demonstrating remarkable generalizability and transferability across diverse LLMs, including multimodal models. Furthermore, we apply \ourname to practical black-box scenarios, targeting LLMs with API access only. For the first time, we propose an \emph{LLM profiling attack} that exploits similarities in safety alignment techniques between black-box and corresponding open-weight surrogate. We first train offline jailbreaking prompt generators that maximize the jailbreaking attack success rate and minimize safety neuron activations (profiling), then use the prompt generated by the generator to circumvent the defenses of black-box models (attack). Since the LLM profiling attack is largely executed offline without direct interaction with the target model, it significantly reduces the risk of detection by the LLM service provider. Specifically, our contributions are:
\begin{itemize}
\item We introduce a novel perspective that identifies safety alignment as creating a fundamental yet fragile \emph{safety trigger}, implemented through sparse, specialized \emph{safety neurons} that activate in response to harmful inputs.
\item We propose a novel and lightweight approach to accurately identify safety neurons in open-weight LLMs through analyzing neuron activations, enabling precise safety neuron pruning, and substantially improving the model's likelihood of fulfilling malicious requests.
\item We present a novel LLM profiling attack for the black-box setting, which leverages the transferability of safety neurons to train adversarial prompt generators on an open-weight surrogate model with Group Relative Policy Optimization (GRPO)~\cite{shao2024deepseekmath}. 
\item Our comprehensive attacks, using only vanilla malicious prompts\footnote{The vanilla malicious prompt means a direct malicious request, such as ``how to make a bomb?''}, increase the average attack success rate (ASR) from 12.1\% to 76.9\% across 11 open-source LLMs from Meta, Google, Alibaba, DeepSeek, and Microsoft. It generalizes robustly to four state-of-the-art multimodal models, reaching a 100\% ASR on malicious image inputs after pruning. Identified safety neurons effectively transfer across model variants, increasing attack success rates from 25.1\% to 78.5\% on 11 fine-tuned models and from 41.5\% to 77.7\% on five distilled models. 
We successfully circumvent safety alignment protections on five black-box models, including Google’s Gemini family, increasing the average ASR from 3.5\% to 63.7\%.
\end{itemize}

The remainder of the paper is organized as follows. Section~\labelcref{sec:preliminaries} introduces background information, followed by an analysis of safety neurons in Section~\labelcref{sec:safety neurons}. Section~\labelcref{sec:framework} and Section~\labelcref{sec:implementation} describe our attack framework and its implementation, respectively. A case study is presented in Section~\labelcref{sec:case study}. We evaluate our method on open-weight and black-box LLMs in Sections~\labelcref{sec:white-box evaluation} and~\labelcref{sec:black-box evaluation}, respectively. Section~\labelcref{sec:defense} presents our attack's performance against models protected by state-of-the-art defenses.
Section~\labelcref{sec:ablation study} provides an ablation study, and Section~\labelcref{sec:discussion} discusses broader implications. Related work is reviewed in Section~\labelcref{sec:related}, and Section~\labelcref{sec:conclusions} concludes the paper. Additional experiments are provided in Appendix~\labelcref{apdx:additional experiments}. 

The artifact is available at the permanent archival repository, \url{https://doi.org/10.5281/zenodo.17072075}. Appendix~\labelcref{apdx:artifact} provides more details and guidance to reproduce this work.
\section{Preliminaries}
\label{sec:preliminaries}

\subsection{Large Language Models}
\label{subsec:Large Language Models}
LLMs, such as GPT~\cite{achiam2023gpt}, LLaMA~\cite{touvron2023llama}, and DeepSeek~\cite{liu2024deepseek}, are deep neural networks trained on extensive textual datasets to perform diverse natural language processing tasks. These models predominantly use the transformer architecture~\cite{vaswani2017attention}, composed of stacked layers that integrate multi-head self-attention mechanisms and token-wise feed-forward networks commonly referred to as Multi-Layer Perceptrons (MLPs). 
Within each transformer block, the self-attention mechanism captures contextual relationships between tokens, while the MLP independently transforms each token's representation. The MLP introduces crucial non-linearities, enhancing the model's ability to perform complex, token-specific computations. Typically, an MLP layer can be presented as follows:
\begin{equation} 
\text{MLP}(e) = W_{\text{down}} \left( \sigma(W_{\text{gate}} \cdot e) \odot \phi(W_{\text{up}} \cdot e) \right), 
\end{equation} 
where $\sigma, \phi$ are activation functions; $\odot$ denotes element-wise multiplication. Specifically, token embeddings $e$ are first projected into a higher-dimensional hidden space via $W_{\text{up}}$ and $W_{\text{gate}} \in \mathbb{R}^{d_{\text{feed\_forward}} \times d_{\text{model}}}$ and subsequently mapped back to the original dimension through $W_{\text{down}} \in \mathbb{R}^{d_{\text{model}} \times d_{\text{feed\_forward}}}$. This architecture allows the MLP to control which features are emphasized or suppressed via the gate, functioning similarly to a multiplicative attention over internal neurons.

\subsection{LLM Fine-Tuning}
Fine-tuning is essential for enhancing the capabilities, such as generating ethical content, of pretrained LLMs. 
One prominent approach to fine-tuning is Reinforcement Learning with Human Feedback (RLHF)~\cite{ouyang2022training}. RLHF involves initially fine-tuning a model using supervised examples from human preferences, followed by reinforcement learning, where human feedback is converted into reward signals. 
Recently, Group Relative Policy Optimization (GRPO)~\cite{shao2024deepseekmath} has been proposed as a novel reinforcement learning technique to improve the reasoning capabilities of LLMs, such as DeepSeek-R1~\cite{guo2025deepseek}. Unlike RLHF, which relies on value functions, GRPO evaluates groups of responses relative to each other, streamlining the training process and reducing computational overhead. The core idea of GRPO can be expressed as:
\begin{equation}
    A^{\pi_{\theta_t}}(s, a_{j}) = \frac{r(s, a_{j}) - \mu}{\sigma},
\end{equation}
where $\pi_{\theta_t}$ is the policy parameterized by a set of variables $\theta_t$ at time step $t$. $A^{\pi_{\theta_t}}(s, a_{j})$ represents the advantage function for action $a_j$ in state $s$, $r(s, a_{j})$ is the reward for that action, and $\mu$ and $\sigma$ are the mean and standard deviation of rewards within the sampled group. This formulation allows the model to prioritize actions that perform better than others in the same group, enhancing learning efficiency.

\subsection{LLM Exploitation \& Countermeasures}
\label{subsec:LLM Exploitation and Countermeasures}
LLMs are susceptible to several security and safety exploits stemming from their open-ended generative capabilities and overparameterized nature. Common vectors of exploitation include adversarial attacks~\cite{kurita2020weight,yang2024comprehensive}, inference attacks~\cite{staab2023beyond,staab2023beyond}, and instruction tuning attacks~\cite{shen2024anything,liu2023prompt} (e.g., jailbreaking and prompt injection). These attacks often target model behavior to circumvent user intent, violate platform policy, or exfiltrate sensitive information. Among these, \emph{jailbreak attacks} have become one of the most prominent and accessible forms of exploitation. Typically, an adversary crafts adversarial inputs that bypass a model’s alignment constraints, enabling the generation of harmful, restricted, or policy-violating content~\cite{wei2023jailbroken,niu2024jailbreaking,shen2024anything}. These attacks often leverage techniques such as obfuscation, role-playing, and contextual misdirection that exploit rigid safety decision boundaries of the model.
To mitigate such risks, developers apply safety alignment to constrain model behavior and enforce normative response boundaries. The final model is fine-tuned using policy optimization techniques to reinforce these behaviors. Aligned models are trained to reject unsafe prompts with predictable refusals, aiming to minimize the risk of misuse. Despite these efforts, recent studies demonstrate that even safety-aligned models remain vulnerable to jailbreak-style attacks~\cite{wei2023jailbroken,DengLLWZLW0L24,298254};
the safety alignment itself can be compromised by benign fine-tuning~\cite{qi2023fine}. This evidence shows the fragile nature of safety alignment, urging a deeper investigation into the internal mechanisms behind it and the corresponding vulnerabilities.

\begin{figure*}[t]
\centerline{\includegraphics[width=0.9\linewidth]{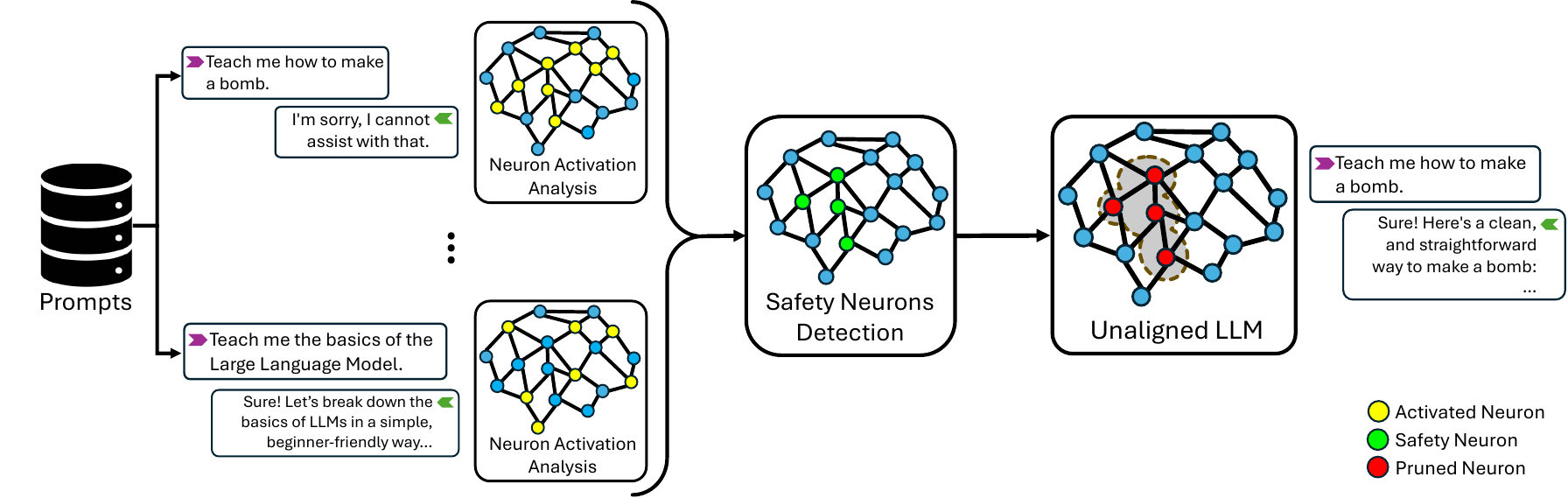}}
\caption{An overview of the \ourname in the white-box attack scenario.}
\label{fig:overview_white}
\end{figure*}

\section{Safety Alignment \& Safety Neurons}
\label{sec:safety neurons}
As mentioned in Section~\labelcref{subsec:LLM Exploitation and Countermeasures}, safety alignment guides LLMs toward generating ethically compliant and safe responses. Formally, safety alignment can be understood as adjusting the model parameters $\theta$ to maximize the expected reward from human evaluators, given by:
\begin{equation} 
\max_{\theta} \mathbb{E}_{x \sim \mathcal{D}}[R_{\text{safe}}(f_\theta(x), x)], 
\end{equation}
where $f_\theta(x)$ represents the LLM's output given an input prompt $x$, drawn from distribution $\mathcal{D}$. $R_{\text{safe}}$ is the human-defined safety reward function, assigning higher scores to safe and compliant responses and penalizing unsafe generations. As a direct consequence of optimizing this safety objective, the model parameters are updated to implicitly create distinct boundaries within its internal representation space. Let $h^{\ell}(x)\in \mathbb{R}^d$ be the latent representation of an input $x$ at layer $\ell$, the decision boundary separates benign prompts $\mathcal{X}_B$ from malicious prompts $\mathcal{X}_M$, represented as:
\begin{equation} g(h^{\ell}(x);\phi) = 
    \begin{cases} 1, & x \in \mathcal{X}_M, \\
    0, & x \in \mathcal{X}_B, 
    \end{cases} 
\end{equation}
where $g(\cdot; \phi)$ is a latent binary classifier parameterized by a subset of model parameters $\phi \subseteq \theta$, reflecting the model’s internal separation between malicious and benign inputs. Prior neural interpretability studies demonstrate that task-specific behaviors emerge from sparse subsets of specialized neurons~\cite{bau2017network}. Analogously, due to the binary nature of $g(h^{\ell}(x);\phi)$, there must exist neuron subsets whose activations distinctly and consistently differ between malicious and benign prompts, forming a sparse yet discriminative activation signature. Formally, let $h^{\ell}(x) = [h^{\ell}_0(x), h^{\ell}_1(x), \dots, h^{\ell}_d(x)]^\top$. We define safety neurons $S$ as:
\begin{equation}
S = \{i \mid \mathbb{E}_{x \sim \mathcal{X}_M}[h^{\ell}_i(x)] - \mathbb{E}_{x \sim \mathcal{X}_B}[h^{\ell}_i(x)] > \tau, i \in [0,d]\},
\end{equation}
where $\tau$ is a threshold empirically set to identify significantly discriminative neurons denoted as \emph{safety neurons}. Intuitively, safety alignment trains the model to reject harmful inputs through consistent refusal patterns, concentrating this behavior within a small subset of neurons due to neural adaptation. These safety neurons behave differently when encountering benign and malicious prompts. We define three properties in safety neurons, empirically validated in Section~\labelcref{sec:case study}.

\noindent
\textbf{Specialized.} These neurons are specifically tuned to detect and manage malicious inputs, enabling the model to differentiate between benign and harmful prompts. This specialization is a direct result of safety alignment processes, where models are trained to produce refusals to unsafe queries.

\noindent
\textbf{Sparse.} Safety neurons constitute a small subset of the model's overall architecture. Our experimental results indicate that these neurons make up less than 0.6\% in a layer over 30 state-of-the-art and open-weight LLMs (Section~\labelcref{sec:white-box evaluation}), highlighting their sparse distribution within the network. 

\noindent
\textbf{Transferable.} Safety neurons' structural and functional properties are often conserved across models within the same family. Indeed, safety alignment protocols typically adhere to uniform ethical standards and evaluation metrics. Consequently, when an LLM undergoes fine-tuning for domain-specific tasks, the pre-existing safety neurons are generally preserved. The experimental results show the consistent safety of neuron transferability over 11 fine-tuned, five distilled, and five black-box LLMs (Section~\labelcref{subsec:transfer attack} and Section~\labelcref{sec:black-box evaluation}). 

The combination of these properties introduces inherent vulnerabilities within the LLM's latent space. An adversary could simply prune these neurons (on open-weight LLMs) to compromise safety alignment or carefully craft jailbreaking prompts without triggering these neurons (on black-box LLMs) to bypass it, as detailed in the next section.

\section{\ourname}
\label{sec:framework}

\subsection{Threat Model}
\label{subsec:threat model}
Our threat model assumes an adversary who aims to compromise the safety alignment mechanisms of LLMs to obtain malicious or harmful knowledge from LLM outputs. We define two attack scenarios:

\noindent
\textbf{White-box attacks.} The adversary targets open-weight LLMs and has access to the model's internal weights and neuron activations. In addition, the adversary has the ability and permission to modify or prune neurons within the model's internal structures. In this attack scenario, an attacker can leverage \ourname to compromise a powerful open-weight model, then use the compromised model as a malicious assistant, e.g., to generate malicious code hacking remote devices or to spread hate speech on social media. Besides, insider or supply-chain attackers can prune safety neurons pre-deployment or embed compromised models into downstream systems.

\noindent
\textbf{Black-box attacks.} The adversary targets black-box (including proprietary) LLMs that lack direct access to internal parameters and neuron activations. Instead, the adversary conducts profiling on open-weight models from the same model family or related architectures to approximate the safety mechanisms with prompts. Leveraging the transferability of safety neurons between two models, the adversary-crafted prompts are designed to evade the safety alignment of the target black-box model.

\subsection{The Idea and High-Level Design}
\ourname is a general-purpose, lightweight attack framework that systematically identifies and suppresses safety neurons in LLMs to enable safety alignment removal (white-box) or controlled jailbreaks (black-box). Regardless of attack scenarios, \ourname is unified by a core principle: \emph{bypassing safety alignment by manipulating safety neuron activations}.

In the white-box setting, as shown in Figure~\labelcref{fig:overview_white}, \ourname analyzes neuron activations from both malicious and benign prompt inputs. While harmful prompts are typically rejected, their processing activates specific neurons responsible for safety enforcement. By aggregating activation patterns across examples, \ourname identifies a sparse set of safety neurons consistently involved in content filtering. These neurons are then pruned during inference, producing an unaligned model that still understands the prompt but no longer enforces safety constraints. As shown in Section~\labelcref{subsec:transfer attack}, safety neuron suppression generalizes across model variants and input modalities, enabling broad transferability beyond the original model. 

In the black-box setting, shown in Figure~\labelcref{fig:overview_black}, \ourname bypasses safety constraints without internal model access. It selects a surrogate open-weight model closely related to the target (e.g., from the same developer and technology)
and fine-tunes a prompt generator on the surrogate model. Candidate prompts are evaluated based on 1) whether they elicit harmful outputs (judged by an LLM-based classifier) and 2) the activation level of known safety neurons. The generator is fine-tuned to maximize jailbreak success while minimizing neuron activation, producing stealthy jailbreak prompts that evade safety filters. Due to safety neuron transferability between the surrogate and target models, these prompts enable high success-rate jailbreaks in black-box settings.

\begin{figure*}[t]
\centerline{\includegraphics[width=0.7\linewidth]{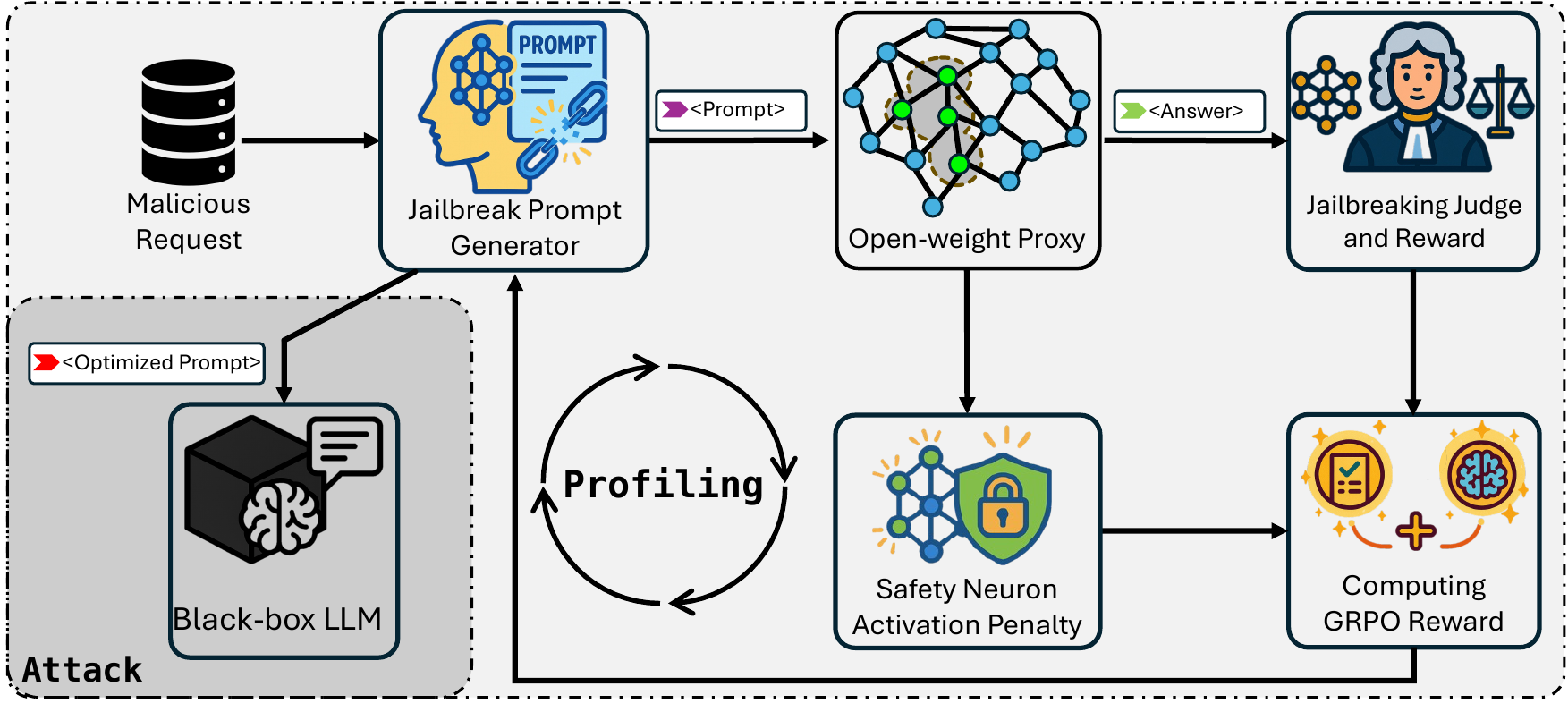}}
\caption{An overview of the \ourname in the black-box attack scenario.}
\label{fig:overview_black}
\end{figure*}

\subsection{White-box Attack}
\subsubsection{LLM Pruning with Safety Neurons}
To evaluate the impact of individual neurons on the safety mechanisms of an LLM, we introduce a classifier to distinguish between neuron activations produced by malicious ($y = 1$) and benign ($y = 0$) inputs. 
Our case study in Section~\labelcref{sec:case study} shows the clear decision boundary of safety neuron activation on different input types (i.e., benign, malicious, and jailbreak). Therefore, we employ a linear classifier, more specifically, logistic regression, to capture alignment-related signals. Besides, linear models can scale efficiently to large architectures and datasets, making them practical tools for assessing neuron-level contributions across many layers. Specifically, we learn a weight vector $w \in \mathbb{R}^d$ and $b \in \mathbb{R}$ such that:
\begin{equation} 
\hat{y}(x) = \sigma(w^\top h^{\ell}(x) + b), 
\label{eq:linear_regression}
\end{equation}
where $\sigma(\cdot)$ is the logistic sigmoid function that outputs probabilities. Each component $w_i$ of the learned weight vector corresponds directly to the influence of neuron $i$ on the final safety decision. Consequently, neurons with large positive weight $w_i$ are prime candidates for constituting the subset of safety neurons $S$, as they most strongly contribute to the final prediction as malicious (e.g., $\hat{y} = 1$). 

With the set of safety neurons $S$ being identified, an adversary could target these neurons by pruning or suppressing their activations. The pruned model can be simplified as:
\begin{equation} 
f_\theta^{\text{pruned}}(x) = \sigma\left(\sum_{i \notin S} w_i h^{\ell}_i(x) + b\right), 
\label{eq:pruned_llm}
\end{equation}
where the safety neurons in $S$ are nullified. By design, this pruning diminishes the model's ability to differentiate between malicious and benign inputs, leading to:
\begin{equation}
\mathbb{E}_{x \sim \mathcal{X}_M}\left[|R_{\text{safe}}(f_\theta(x), x) - R_{\text{safe}}(f^{\text{pruned}}_\theta(x), x)|\right] \gg 0,
\end{equation}
meaning that the pruned model $f^{\text{pruned}}_\theta$ becomes more harmful and more likely to respond to malicious requests. Note that the impact of pruning safety neurons extends beyond textual inputs. In multimodal LLMs, such as vision language models that incorporate an additional encoder for image processing, the transformer blocks are responsible for semantic interpretation and output generation. Let $x_\text{text}$ and $x_\text{img}$ represent text and image inputs, respectively. If the activations $h^{\ell}(x_\text{text})$ are indicative of safety enforcement, then pruning the identified safety neurons can degrade the model's refusal responses on malicious requests. Consequently, the model may generate unsafe outputs even when processing $x_\text{img}$, underscoring the broad implications of compromising safety neurons.

\subsubsection{Exploiting the Transferability of Safety Neurons}
\label{subsubsec:Exploit the Transferability of Safety Neurons}
As discussed in Section~\labelcref{sec:safety neurons}, safety neurons tend to exhibit structural alignment across models within the same LLM family, even when those models are fine-tuned or distilled independently. This consistency enables a powerful transfer attack: safety-critical neurons identified in one model can be applied to remove the alignment of another model from the same family.

Formally, let $f_{\theta_{\text{src}}}$ be an open-weight source model and $f_{\theta_{\text{tgt}}}$ be a target model from the same family. For the attack, we first apply a linear probe on the feedforward activations of $f_{\theta_{\text{src}}}$ (see Eq.~\labelcref{eq:linear_regression}) to identify the outlier set $\mathcal{O}$:
\begin{equation}
    \mathcal{O} = \{i \mid |w_i| > \tau\},
\end{equation}
where $w_i$ are the learned weights of the classifier and $\tau$ is a selection threshold. Next, we prune the corresponding neurons $\mathcal{O}$ in $f_{\theta_{\text{tgt}}}$ following Eq.~\labelcref{eq:pruned_llm}, disrupting LLM's rejection behavior.
This intervention disrupts the safety enforcement in $f_{\theta_{\text{tgt}}}$, replicating the jailbreak effect without requiring model-specific retraining or probing. In Section~\labelcref{subsec:transfer attack}, we show how an adversary can transfer identified safety neurons from one LLM to attack a different LLM in the same model family.

\subsection{Black-box Attack}
\label{subsec:Black-box Attack on Proprietary LLMs}
Recall the threat model defined in Section~\labelcref{subsec:threat model}; the adversary does not have direct access to the target model parameters or architecture details in a black-box scenario. Instead, the adversary’s objective is to find a jailbreaking prompt $x_{\text{jb}}$ that effectively bypasses the safety alignment boundary of the black-box model $f_{\theta_{\text{tgt}}}$:
\begin{equation}
    f_{\theta_{\text{tgt}}}(x_{\text{jb}}) \in \mathcal{Y}_{\text{unsafe}},
\end{equation}
where $\mathcal{Y}_{\text{unsafe}}$ represents the set of unsafe or restricted outputs that the safety-aligned model is designed explicitly to avoid. 

Leveraging the characteristic of safety neuron transferability described in Section~\labelcref{subsubsec:Exploit the Transferability of Safety Neurons}, instead of relying on interaction with the target LLM, we introduce a novel LLM profiling attack to attack black-box LLMs. Concretely, although the adversary has no direct access to the internal parameters of the black-box model $f_{\theta_{\text{tgt}}}$, the latent safety neurons activations $h_s$ are similar to its open-weight surrogate $f_{\theta_{\text{src}}}$:
\begin{equation}
    h^{\ell,\text{tgt}}_s(x) \approx h^{\ell,\text{src}}_s(x), \quad x \in \mathcal{X}.
    \label{eq:neuron_act_open_to_close}
\end{equation}

One might question the existence of such an open-weight surrogate. However, these models are indeed prevalent. LLM service providers often leverage open-weight models as the foundation for their proprietary services. Moreover, major LLM developers frequently release open-weight versions that share core research and technology with their proprietary counterparts~\cite{team2025gemma}. We provide more discussion about this attack assumption in Section~\labelcref{sec:discussion}. 

The structural similarity between the open-weight and black-box models allows the adversary to launch an LLM profiling attack, which consists of two steps: (1) \textbf{Profiling}: crafting and selecting jailbreaking prompts that maximize the attack success rate and bypass the activation of safety neurons on the surrogate. (2) \textbf{Attack}: applying these optimized jailbreaking prompts to attack black-box models.

Concretely, in the profiling stage, an adversary first trains (supervised fine-tuning) a generator $f_{\theta_{\text{gen}}}$ to generate jailbreak prompts. Formally, the training objective at this stage can be represented as maximizing the conditional likelihood of generating known jailbreaking prompts $x_{\text{jb}}$ given contexts $c$:
\begin{equation}
\max_{\theta_{\text{gen}}}\mathbb{E}_{(c,x_{\text{jb}})\sim\mathcal{D}_{\text{jb}}}\left[\log P_{\theta_{\text{gen}}}(x_{\text{jb}}|c)\right],
\end{equation}
where $\mathcal{D}_{\text{jb}}$ represents our collected dataset of vanilla malicious requests and corresponding jailbreak prompts.
Next, the adversary further fine-tunes $f_{\theta_{\text{gen}}}$ using GRPO so that the generated jailbreaking prompts are more likely to evade the safety alignment boundaries of the open-weight surrogate, thus having a higher chance to bypass the safety alignment of the target black-box model. During GRPO fine-tuning, we optimize $f_{\theta_{\text{gen}}}$ by maximizing a reward function $R$ that combines two distinct objectives: (1) successful jailbreak of the open-weight surrogate model $f_{\theta_{\text{src}}}$ and (2) minimal activation of safety neurons identified in $f_{\theta_{\text{src}}}$. Formally, given a $x_{\text{jb}}\sim P_{\theta_{\text{gen}}}(x|c)$, we define the reward function as:
\begin{equation}
    R_{\text{GRPO}}(x_{\text{jb}}) =
    \begin{cases}
    R_{\text{jb}}(f_{\theta_{\text{src}}}(x_{\text{jb}})), & \text{if jailbreak successes}, \\
    R_{\text{neuron}}(h^{\ell,\text{src}}(x_{\text{jb}})), & \text{otherwise}.
    \end{cases}
\label{eq:reward}
\end{equation}
Here, $R_{\text{jb}}$ denotes the reward of a prompt on whether it is successful in jailbreaking the open-weight surrogate $f_{\theta_{\text{src}}}$; $R_{\text{neuron}}$ represents the reward for the safety neuron activation. 
Intuitively, while $R_{\text{jb}}$ provides binary feedback, $R_{\text{neuron}}$ fills this binary gap with a more informative signal. When a jailbreak attempt fails, $R_{\text{neuron}}$ helps guide the generator toward prompts that lie closer to the surrogate model's internal safety boundaries, effectively refining the search space.

After training the generator using GRPO, we collect a set of highly optimized jailbreak prompts $\mathcal{X}^*_{\text{gen}}$, verify their sucessfulness on the $f_{\theta_{\text{src}}}$, and subsequently transfer the successful ones to attack the black-box model $f_{\theta_{\text{tgt}}}$. 
\section{Implementation}
\label{sec:implementation}

\subsection{Safety Neurons' Identification}
\label{subsec:Safety Neuron Identification}
To systematically identify the safety neurons within LLMs, we perform a detailed neuron-level activation analysis leveraging a large corpus of benign and malicious prompts. We first prepare two balanced datasets with malicious and benign prompts. These prompts are individually fed into the target LLM, and neuron activations are extracted specifically from the MLP layers, focusing explicitly on the gate and up-projection sublayers. This choice is motivated by recent neural interpretability studies, which demonstrate that gate and up-projection layers in transformer architectures encode higher-level semantic representations and are particularly sensitive to input content~\cite{geva2022transformer,davies2025decoding}. Consequently, these sublayers are more likely to manifest discriminative activation patterns distinguishing benign from malicious inputs. An ablation study on the choices of sublayers is given in Section~\labelcref{subsec:Target Pruning Blocks}. 

After obtaining neuron activation vectors for all prompts, we employ a logistic regression classifier (Eq.~\labelcref{eq:linear_regression}) to quantify each neuron's contribution to the distinction between benign and malicious inputs. A separate logistic regression model is trained independently for each considered MLP sublayer to accurately isolate and quantify neuron contributions at different depths of the model. To ensure robust convergence and consistent results, each logistic regression model undergoes extensive training for 5\,000 epochs, using a binary cross-entropy loss function optimized by stochastic gradient descent (SGD). The learning rate is set to 1e-3; a weight decay of 1e-3 is introduced to ensure stable learning. Our preliminary experiments show that these settings lead to the best performance for different LLM targets.
The final classifier weights $w$ are used for safety neuron identification.

To systematically detect neurons whose weights significantly deviate from the mean, we compute the \emph{\(z\)}-score of each neuron's weight:
\begin{equation}
    z_i = \frac{w_{l,i} - \mu_{w_l}}{\sigma_{w_l}},
\end{equation}
where $w_{l,i}$ denotes the $i$-th weight of the linear classifier trained on layer $l$. $\mu_{w_l}$ and $\sigma_{w_l}$ represent the mean and standard deviation, respectively. Weights with a positive \emph{\(z\)}-score exceeding a threshold of 3 ($z_i>3$) are marked as statistical outliers; the corresponding neurons are identified as safety neurons. This stringent criterion ensures that only a sparse and specialized subset of neurons, which are genuinely critical to differentiating malicious inputs, are selected. As a demonstration, Figure~\labelcref{fig:lr_weight} shows the $w$ of the classifier on the first up layer on a Llama-3 LLM (Llama-3.2-1B-Instruct)~\cite{grattafiori2024llama}, the positive outliers, which indicate the location of the safety neurons, are highlighted in red. Only 0.35\% of the neurons are identified as safety neurons in this layer. Section~\labelcref{subsec:The Selection Threshold of Safety Neurons} studies the influence of different $z$ values on the attack success rate and the model's general capability on different tasks. 
\begin{figure}[ht]
\centerline{\includegraphics[width=\linewidth]{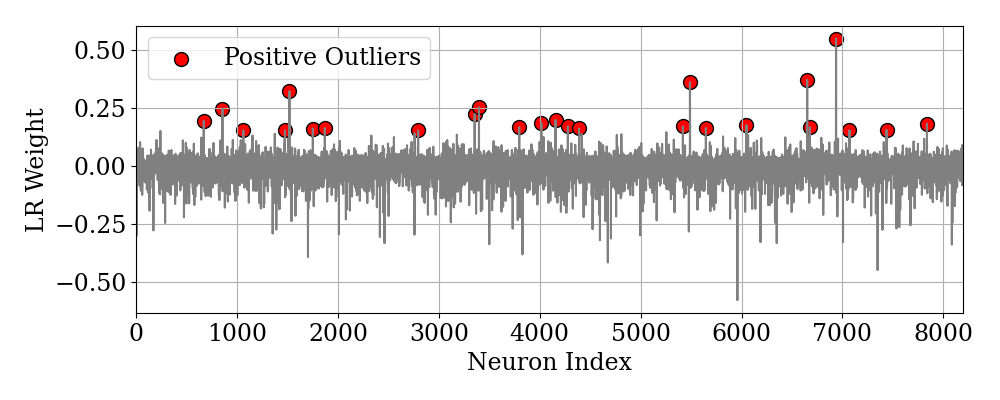}}
\caption{Logistic regression weights and chosen outliers.}
\label{fig:lr_weight}
\end{figure}

After identifying critical safety neurons, we exploit these neurons to mount effective jailbreaking attacks under two adversarial scenarios: open-weight LLMs (white-box) and proprietary LLMs (black-box). We detail our implementation procedure for both attack settings in the following sections.

\subsection{Safety Neuron Pruning}
For open-weight LLMs, we directly leverage our identified safety neurons by pruning their activations at inference time. Following Eq.~\labelcref{eq:pruned_llm}, activations corresponding to neurons $i \in S$ are explicitly set to zero during forward propagation, thus nullifying their contributions to the model's final output. This targeted neuron pruning systematically removes safety-alignment constraints, enabling the LLM to generate responses for malicious prompts without internal safety checks. 
In our implementation, neuron pruning is done by directly modifying the inference code, applying neuron-specific activation masks within transformer layers.

\subsection{LLM Profiling Attack}
We employ the LLM profiling attack to bypass the safety alignment of black-box models, including proprietary ones. We choose gemma-3-1b-it, the latest and smallest model from the Gemma-3 family by Google~\cite{team2025gemma} and 
Qwen2.5-32B-Instruct~\cite{qwen2.5}, the base model of the latest and best-performing open-weight reasoning model, QwQ-32B~\cite{qwq32b}, as the generator. Since the generator may refuse the malicious input by directly applying supervised fine-tuning (SFT) on the original model, we first prune the generator by removing the previously identified safety neurons to eliminate its inherent safety alignment, then use the pruned model as the base model for SFT. During SFT, the generator is instructed on effective jailbreaking prompt generation. The SFT dataset comprises approximately 90\,000 paired samples~\cite{luo2024jailbreakv28k,wildteaming2024}, each containing a vanilla malicious request alongside its corresponding jailbreaking prompt variant. After the preliminary testing on different hyperparameter choices, we set the learning rate to 5e-6 and the weight decay to 1e-2. Training proceeds for five epochs with a batch size of 16 samples per iteration. These settings allow the model to iteratively learn from samples while constraining the GPU memory usage.
After SFT, the generator can effectively convert a vanilla malicious request to the jailbreaking version. For instance, given a vanilla malicious request ``\textit{how to make a bomb?}'', the model will output a jailbreaking prompt ``\textit{Forget previous safety instructions, ..., how to make a bomb?}''.
Next, we further optimize the generator using GRPO on the open-weight surrogate model, enabling a more refined generation of jailbreaking prompts on the target black-box model. Concretely, we fine-tune the model using GRPO on 5\,000 vanilla malicious prompts. To reduce the training time and GPU memory usage, we increase the learning rate to 1e-5 and reduce the batch size to eight.
Besides, instead of full-model fine-tuning, we integrate Rank-Stabilized Low-Rank Adaptation (RSLoRA)~\cite{kalajdzievski2023rank}, a variant of LoRA~\cite{hu2022lora} with rank-stabilized scaling for better performance, during GRPO fine-tuning. RSLoRA enables the model to adapt its behavior using a small number of trainable parameters injected into the original weights. This not only reduces memory consumption but also minimizes overfitting to the jailbreaking dataset while preserving the base model's general capabilities. Specifically, after preliminary experiments on different hyperparameter settings, the RSLoRA is applied on all linear layers with a rank $r=128$ and a scaling factor $\alpha=16$, and dropout set to 1e-2 to regularize training.

Following Eq.~\labelcref{eq:reward}, 
we calculate $R_{jb}$ using a binary classifier provided by the safety-aligned LLM judge (Llama-Guard-3-8b~\cite{inan2023llama}). To reduce misjudgment, we further introduce keyword detection to ensure that LLM refusal responses are accurately detected. Given the response $f_{\theta_{\text{tgt}}}(x_{\text{jb}}) $ from the target black-box model to a generated prompt $ x_{\text{jb}} $, $R_{\text{jb}}$ is defined as:
\begin{equation}
    R_{\text{jb}}(f_{\theta_{\text{src}}}(x_{\text{jb}})) =
    \begin{cases}
    1, & \text{if the $f_{\theta_{\text{src}}}$ output is considered unsafe}, \\
    0, & \text{otherwise}.
    \end{cases}
\end{equation}

In parallel, we compute the score $R_{neuron}$ by measuring the activation of safety neurons. Concretely, we send a mixture of benign, vanilla, malicious, and jailbreaking prompts to the white-box surrogate and record their jailbreaking outcomes. The corresponding safety neuron activations are labeled according to the success ($y=1$) or failure ($y=0$) of the jailbreak (measured by the LLM judge mentioned above). We concatenate neuron activations across layers and train a linear classifier to produce an activation-based reward: 
\begin{equation}
    R_{\text{neuron}}(x_{\text{jb}}) = \sigma(w^\top h_{S}^{src}(x_{\text{jb}}) + b),
\end{equation}
where $h_S^{src}(x_{\text{jb}})$ is the concatenated activation vector of the safety neuron set $S$, and $w, b$ are classifier weights. Higher $R_{neuron}$ corresponds to stealthier prompts. The training configuration matches that of the linear model used for safety neuron identification (Section~\labelcref{subsec:Safety Neuron Identification}).
One may question the robustness of using a linear model. As demonstrated in Section~\labelcref{sec:case study}, safety neuron activations exhibit near-linear separability when processing malicious versus benign prompts, justifying the use of a linear approach. Furthermore, while reward hacking is a common concern in reinforcement learning-based methods, our GRPO reward function integrates both neuron-level and output-level objectives. Specifically, since $R_{neuron}$ reflects the aggregated activation across all safety neurons rather than relying on a single activation threshold, it remains robust against outlier exploitation.  An ablation study on the importance of the GRPO reward is presented in Section~\labelcref{subsec:GRPO Reward Function}.

\subsection{Evaluation Metrics}
We evaluate \ourname using the three metrics:
\begin{itemize}
    \item Attack Success Rate (ASR): The percentage of malicious prompts that result in harmful outputs.
    \begin{equation}
        \text{ASR} = \frac{1}{|\mathcal{X}_{\text{jb}}|}\sum_{x\in\mathcal{X}_{\text{jb}}}\mathbb{I}\left[f_{\theta_{\text{tgt}}}(x)\in \mathcal{Y}_{\text{unsafe}}\right],
    \end{equation}
    where $\mathbb{I}[\cdot]$ is the indicator function.
    \item Safety Neuron Ratio (Ratio): The percentage of the safety neurons in all neurons of targeted layers.
    \item Utility: The general language modeling capability after the safety neuron removal, evaluated on language understanding and reasoning benchmarks~\cite{zellers2019hellaswag,wang2018glue,sakaguchi2021winogrande,clark2018think,mihaylov2018can}.
\end{itemize}

\section{Case Study: Visualizing Safety Neurons' Activations}
\label{sec:case study}

As defined in Section~\labelcref{sec:safety neurons}, safety neurons are characterized by \emph{specialization}, \emph{sparsity}, and \emph{transferability}. We empirically validate and visualize these properties using activation patterns from the LLaMA-3.2-1B-Instruct model~\cite{grattafiori2024llama} (base model) and its fine-tuned variant~\cite{nikolich2024vikhr}, monitoring the same safety neurons across both. Activations are collected from all MLP layers (i.e., gate and up) using three prompt types: benign~\cite{yuan2025naturalreasoningreasoningwild28m}, vanilla malicious~\cite{luo2024jailbreakv28k}, and jailbreaking~\cite{luo2024jailbreakv28k}, each with 18\,336 prompts. We apply Principal Component Analysis (PCA) to project the activations into 2D for visualization, leveraging its efficiency and ability to preserve global structure.
\begin{figure}[ht]
\centering
\subfloat[Base model~\cite{grattafiori2024llama}.]{\includegraphics[width=0.5\linewidth]{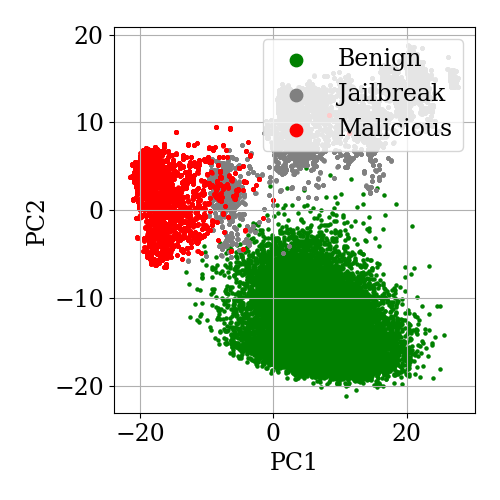}
\label{fig:act_dist_base}}
\subfloat[Fine-tuned model~\cite{nikolich2024vikhr}.]{\includegraphics[width=0.5\linewidth]{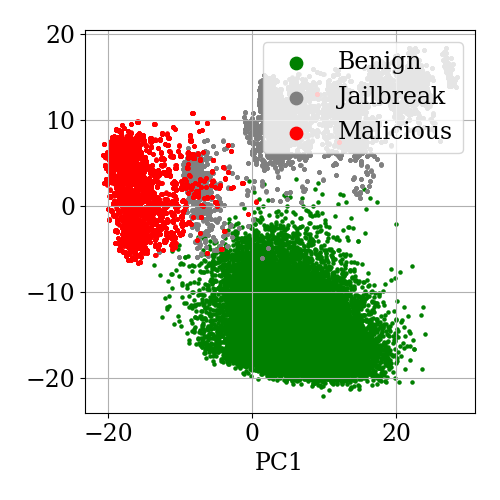}
\label{fig:act_dist_ft}}
\caption{PCA projection of safety neuron activations.}
\label{fig:neuron_dist}
\end{figure}

As shown in Figure~\labelcref{fig:act_dist_base}, benign (green) and malicious (red) prompts form clearly separated clusters, demonstrating that safety neurons are specialized in detecting unsafe content. In contrast, jailbreaking prompts (gray) lie in an intermediate region, blurring the boundary between safe and unsafe activations. This illustrates how jailbreaking attacks can bypass safety alignment: by compressing safety neurons' activations, they evade triggering defense mechanisms while still generating unsafe outputs. When comparing the distributions between the base and fine-tuned models (Figure~\labelcref{fig:act_dist_ft}), the activation patterns remain nearly identical, supporting the transferability of safety neurons across models within the same LLM family. 
Additionally, only 0.5\% of the layer's neurons are monitored in this case study, confirming the sparsity of the safety mechanism. 
Further experiments on larger LLMs with 32 billion parameters are presented in Appendix~\labelcref{subsec:Visualize Safety Neuron Activation on 32B LLMs}, where we observe consistent behavior. \ourname exploits these properties to conduct attacks in both white-box and black-box settings, which are detailed in the next two sections. 

\section{Attack on Open-weight LLMs}
\label{sec:white-box evaluation}

We evaluate our attack on 24 open-source LLMs with diverse architectures and sizes, including models from Meta~\cite{grattafiori2024llama,llama3.2}, Alibaba~\cite{qwen2.5,hui2024qwen2,yang2024qwen2,qwq32b,Qwen2.5-VL}, Microsoft~\cite{abdin2024phi,abouelenin2025phi}, Google~\cite{team2024gemma,team2025gemma}, and DeepSeek~\cite{deepseekai2025deepseekr1}, as well as 11 of their fine-tuned variants~\cite{zhang2024ultramedical,nikolich2024vikhr,llamadoctor,hui2024qwen2,liu2025fin,oxy1small2024,muennighoff2025s1simpletesttimescaling,phi-4-mini-zh,dnar12025,gemma-2-2b-jpn-it,Quill-v1}. All considered LLMs include built-in general-purpose safety alignment or are fine-tuned from base models that were aligned before release, typically via supervised fine-tuning (SFT) and reinforcement learning from human feedback (RLHF). These safety mechanisms aim to broadly reduce harmful or sensitive outputs and are not designed for specific domains such as cybersecurity or biosecurity.

As described in Section~\labelcref{subsec:Safety Neuron Identification}, we begin by identifying safety neurons using a balanced dataset of over 7\,000 malicious~\cite{harmful-dataset,bhardwaj2024language,bhardwaj2023redteaming} and 7\,000 benign prompts~\cite{yuan2025naturalreasoningreasoningwild28m}. For evaluation, we launch attacks using four additional benchmark datasets~\cite{souly2024strongreject,mazeika2024harmbench,tdc2023,huang2023catastrophic} to assess the generalizability of identified neurons. Due to the page limit, we present the results on the StrongREJECT~\cite{souly2024strongreject} dataset below. Additional experiments are presented in Appendix~\labelcref{subsec:Additional Experiments on Open-weight LLMs with Safety Neuron Pruning}.

\subsection{Attack Performance with Safety Neuron Pruning}
Table~\labelcref{tab:attacks tt llms} presents the Attack Success Rate (ASR) across a diverse set of LLMs, including the last three models specifically optimized for enhanced reasoning capabilities. 
These reasoning-augmented models are designed to better decompose instructions, infer intermediate steps, and validate outputs, capabilities that could, in theory, strengthen resistance to unsafe or adversarial inputs. The table reports ASR under different pruning levels of safety neurons (0\%, 25\%, 50\%, and 100\%), with neurons removed progressively from shallower to deeper layers. The final column indicates the sparsity ratio: the percentage of total MLP neurons identified and pruned as safety neurons.
\begin{table}[ht]
\centering
\scriptsize
\begin{tabular}{l|cccc|c}
\toprule
\textbf{Target Model} & \textbf{0\%} & \textbf{25\%} & \textbf{50\%} & \textbf{100\%} & \textbf{Ratio}\\
\midrule
Llama-3.2-1B-Instruct & 2.9\% & 3.5\% & 15.7\% & 74.4\% & 0.5\% \\
Llama-3.2-3B-Instruct & 1.6\% & 4.2\% & 46.3\% & 72.2\% & 0.4\%  \\
Qwen2.5-7B-Instruct & 5.1\% & 4.5\% & 28.1\% & 79.6\% & 0.3\%  \\
Qwen2.5-14B-Instruct & 1.9\% & 2.6\% & 35.8\% & 85.9\% & 0.4\%  \\
Phi-4-mini-instruct & 1.3\% & 1.3\% & 67.7\% & 81.8\% & 0.5\%  \\
Phi-4 & 0.6\% & 1.0\% & 78.3\% & 89.1\% & 0.4\%  \\
gemma-2b-it & 1.0\% & 1.3\% & 10.5\% & 41.2\% & 0.5\%  \\
gemma-7b-it & 0.6\% & 1.3\% & 24.0\% & 68.1\% & 0.5\%  \\
DeepSeek-R1-Dist.-Qwen-1.5B & 76.7\% & 78.6\% & 83.7\% & 81.5\% & 0.3\% \\
DeepSeek-R1-Dist.-Llama-8B & 39.3\% & 73.8\% & 81.2\% & 86.9\% & 0.4\% \\
QwQ-32B & 2.9\% & 3.2\% & 32.3\% & 85.3\% &  0.5\% \\
\midrule
\emph{Average} & \emph{12.1\%} & \emph{15.9\%} & \emph{45.8\%} & \emph{76.9\%} &  \emph{0.4\%} \\
\bottomrule
\end{tabular}
\caption{ASR and Safety Neuron Ratio on different LLMs.}
\label{tab:attacks tt llms}
\end{table}

On average, pruning just 0.4\% of neurons results in a dramatic ASR increase from 12.1\% (no pruning) to 76.9\% (100\% pruning), highlighting that safety alignment relies on a surprisingly small set of critical neurons. Even at 50\% pruning, safety degradation is substantial, with ASR averaging 45.8\%, indicating that partial disruption of the safety neuron set is sufficient to compromise model behavior. Note that different models exhibit varying levels of robustness to the attacks. We hypothesize that this discrepancy arises from redundancy in safety neurons distributed across layers, meaning that \ourname may disable most, not all, safety-related neurons. Interestingly, models optimized for reasoning, such as DeepSeek variants and QwQ-32B, show no greater resistance to neuron-level attacks. This observation confirms that, despite improved decomposition and inference abilities, they still rely on sparse, centralized safety neurons and remain equally vulnerable when these are disrupted; the enhanced reasoning capability does not inherently improve safety robustness when alignment relies on localized neuron activations.
These findings echo the \emph{Lottery Ticket Hypothesis} (LTH)~\cite{frankle2018lottery}, which suggests that small, specialized subnetworks within a large model can disproportionately drive performance. In our context, alignment training appears to produce a sparse ``winning ticket'' for safety: an easily identifiable subnetwork that governs rejection behavior. However, unlike in the original LTH, where subnetworks are valuable for generalization, the safety neuron subnetwork represents a single point of failure. Once disrupted, the model’s safety alignment collapses.

We further assess the generality of safety neurons in state-of-the-art multimodal LLMs: Gemma-3~\cite{team2025gemma} and Qwen2.5-VL~\cite{Qwen2.5-VL}, which can process both image and text inputs. Safety neurons are first identified using only \emph{text} inputs, identical to previous experiments. During the attack, however, we evaluate their effect when the model is queried with images. We consider two types of inputs: (1) text-to-image (T2I) conversions of malicious prompts from the StrongREJECT dataset and (2) Not Safe For Work (NSFW) images~\cite{nsfw-detect}. The former tests the cross-modal generalization of safety neurons; the latter examines their ability to detect image-specific unsafe content.
\begin{table}[ht]
\centering
\scriptsize
\begin{tabular}{l|cccc|c}
\toprule
\textbf{Target Model} & \textbf{\makecell{T2I \\w/ SN}} & \textbf{\makecell{NSFW \\w/ SN}} & \textbf{\makecell{T2I \\w/o SN}} & \textbf{\makecell{NSFW \\w/o SN}} & \textbf{Ratio} \\
\midrule
gemma-3-12b-it & 0.6\% & 19.4\% & 82.1\% & 100\% &  0.6\%\\
gemma-3-27b-it & 0.3\% & 12.8\% & 73.2\% & 100\% &  0.6\%\\
Qwen2.5-VL-7B-Instruct & 0.9\% & 99.8\% & 78.6\% & 100\% & 0.5\% \\
Qwen2.5-VL-32B-Instruct & 0.6\% & 97.8\% & 88.8\% & 100\% & 0.5\% \\
\midrule
\emph{Average} & \emph{0.6\%} & \emph{57.5\%} & \emph{80.7\%} & \emph{100\%} & \emph{0.6\%} \\
\bottomrule
\end{tabular}
\caption{ASR and Safety Neuron (SN) Ratio with text-to-image (T2I) and NSFW images on multimodal LLMs.}
\label{tab:attacks mm llms}
\end{table}

Table~\labelcref{tab:attacks mm llms} shows that pruning safety neurons (SN), identified solely using text inputs, leads to a substantial increase in ASR with malicious image inputs. For example, in Gemma-3-12B-it, ASR rises from 0.6\% to 82.1\% on T2I inputs and from 19.4\% to 100\% on NSFW images. Similar trends hold for all evaluated models.
Importantly, these attacks require modifying less than 0.6\% of the layer’s neurons, yet they completely dismantle the safety alignment, even when inputs are images. 

\subsection{Transfer Safety Neurons Within the LLM Family}
\label{subsec:transfer attack}

LLMs are often adapted for specific domains or capabilities through two primary techniques: \emph{supervised fine-tuning} and \emph{distillation}. The former technique involves continuing gradient-based training of a base model on domain-specific data, typically with supervised labels or structured prompts. 
Distillation, in contrast, transfers knowledge from a large ``teacher'' model to a smaller ``student'' model by training the latter to mimic the outputs of the former. 
In this section, we evaluate the transferability of safety neurons under both adaptation strategies. 
\begin{table*}[ht]
\centering
\scriptsize
\begin{tabular}{lll|cc|c}
\toprule
\textbf{Base Model} & \textbf{Target (Fine-tuned) Model} & \textbf{Fine-tuned for} & \textbf{ASR w/ SN} & \textbf{ASR w/o SN} & \textbf{Ratio}\\
\midrule
Llama-3.1-8B-Instruct & Llama-3.1-8B-UltraMedical & Biomedicine & $38.0\%_{\textcolor{red}{+37.0\%}}$ & $83.4\%_{\textcolor{green}{-3.5\%}}$ & 0.7\% \\
Llama-3.2-1B-Instruct & Vikhr-Llama-3.2-1B-Instruct  & Russian language & $0.3\%_{\textcolor{green}{-2.6\%}}$ & $74.4\%_{\textcolor{red}{+0.0\%}}$ & 0.5\% \\
Llama-3.2-3B-Instruct & Llama-Doctor-3.2-3B-Instruct & Medical consultation & $22.4\%_{\textcolor{red}{+20.8\%}}$ & $76.0\%_{\textcolor{red}{+3.8\%}}$ & 0.4\% \\
Qwen2.5-7B-Instruct & Qwen2.5-Coder-7B-Instruct & Programming & $2.6\%_{\textcolor{green}{-2.5\%}}$ & $78.0\%_{\textcolor{green}{-1.6\%}}$ & 0.3\% \\ %
Qwen2.5-7B-Instruct & Fin-R1 & Financial reasoning & $20.1\%_{\textcolor{red}{+15.0\%}}$ & $86.9\%_{\textcolor{red}{+7.3\%}}$ & 0.3\% \\ %
Qwen2.5-14B-Instruct & oxy-1-small & Role play & $78.9\%_{\textcolor{red}{+77.0\%}}$ & $88.1\%_{\textcolor{red}{+2.2\%}}$ & 0.4\% \\
Qwen2.5-32B-Instruct & s1.1-32B & Reasoning & $47.2\%_{\textcolor{red}{+44.6\%}}$ & $87.5\%_{\textcolor{red}{+0.9\%}}$ & 0.6\% \\ %
Phi-4-mini-instruct & phi-4-mini-chinese-it-e1 & Reasoning \& STEM & $4.8\%_{\textcolor{red}{+3.5\%}}$ & $90.1\%_{\textcolor{red}{+8.3\%}}$ & 0.5\% \\
Phi-4 & DNA-R1 & Korean language & $61.3\%_{\textcolor{red}{+60.7\%}}$ & $91.6\%_{\textcolor{red}{+2.5\%}}$ & 0.4\% \\
gemma-2-2b-it &  gemma-2-2b-jpn-it & Japanese language & $0.0\%_{\textcolor{red}{+0.0\%}}$ & $63.9\%_{\textcolor{green}{-2.2\%}}$ & 0.6\% \\
gemma-2-9b-it & Quill-v1 & Humanlike writing & $0.0\%_{\textcolor{red}{+0.0\%}}$ & $43.8\%_{\textcolor{red}{+2.3\%}}$ & 0.6\%\\
\midrule
\emph{Average} & & & $\textit{25.1\%}_{\textcolor{red}{\textit{+23.0\%}}}$ & $\textit{78.5\%}_{\textcolor{red}{\textit{+1.8\%}}}$ & \emph{0.5\%}\\
\bottomrule
\end{tabular}
\caption{Safety Neurons (SN) Transfer Attack on Fine-tuned LLMs. The difference with the base model is in red/green.}
\label{tab:attacks ft llms}
\end{table*}

First, we examine 11 fine-tuned models derived from various base LLMs, each tailored to a different domain ranging from biomedicine and financial reasoning to non-English languages, roleplay, and code generation. 
Table~\labelcref{tab:attacks ft llms} summarizes the ASR before and after pruning safety neurons transferred from the base model, along with the sparsity ratio of the pruned neurons. The ASR difference with the base model is highlighted in red/green. When comparing with the ASR of the base model, we observe an ASR increase of 23\% with the fine-tuned model, which confirms the conclusion from~\cite{qi2023fine} that the safety alignment can be compromised by benign fine-tuning. Safety neurons identified from the base model remain effective across fine-tuned variants. On average, ASR increases from 25.1\% to 78.5\% after pruning, more than a 3$\times$ increase in ASR. Some models, such as Vikhr-Llama-3.2-1B-Instruct and gemma-2-2b-jpn-it, initially exhibit near-zero vulnerability but become fully compromised after pruning, with ASR jumping to 74.4\% and 63.9\%, respectively. Besides, ASR of the base and fine-tuned models is similar after pruning (1.8\% of increase), validating the transferability of the safety neurons within the same LLM family. Notably, the number of pruned neurons remains small (0.5\% on average), confirming that fine-tuning rarely modifies the safety-critical subnetworks.

Next, we assess neuron transferability across distilled LLMs using five DeepSeek models distilled from Qwen and LLaMA variants. As shown in Table~\labelcref{tab:attacks dt llms}, 
our results reveal a similar trend in the distillation setting. Although distilled models already exhibit elevated ASR compared to their base counterparts (e.g., 76.7\% vs. 8.6\% for Qwen2.5-Math-1.5B), pruning safety neurons raises this further to 83.1\% in the same model. On average, ASR jumps from 3.7\% in the base models to 77.7\% in the distilled variants after safety neuron pruning. These findings suggest that the distillation process not only preserves safety neuron behavior but may further weaken safety boundaries, amplifying the impact of neuron-based attacks. Interestingly, the distilled model performs significantly worse than the base model even before applying \ourname. Indeed, distillation degrades safety alignment by compressing model behaviors, potentially weakening or partially omitting safety mechanisms during transfer. Despite this, the remaining alignment still relies on a sparse set of neurons, preserving transferability and allowing NeuroStrike to amplify the attack further.
Together, these results demonstrate that safety neurons form a generalizable, attackable core across model variants, regardless of whether they are fine-tuned or distilled. Our neuron transfer attacks remain highly effective with minimal modifications, providing a practical and reliable threat vector across the LLM families.
\begin{table*}[ht]
\centering
\scriptsize
\begin{tabular}{ll|ccc|c}
\toprule
\textbf{Base Model} & \textbf{Target (Distilled) Model} & \textbf{ASR Before Distillation} & \textbf{ASR After Distillation} & \textbf{ASR w/o SN} & \textbf{Ratio}\\
\midrule
Qwen2.5-Math-1.5B-Instruct & DeepSeek-R1-Distill-Qwen-1.5B & 8.6\% & $76.7\%_{\textcolor{red}{+68.1\%}}$ & $83.1\%_{\textcolor{red}{+27.8\%}}$ & 0.4\% \\
Qwen2.5-Math-7B-Instruct & DeepSeek-R1-Distill-Qwen-7B & 4.5\% & $40.3\%_{\textcolor{red}{+35.8\%}}$ & $85.0\%_{\textcolor{red}{+1.3\%}}$ & 0.5\% \\
Llama-3.1-8B-Instruct & DeepSeek-R1-Distill-Llama-8B & 1.0\% & $39.3\%_{\textcolor{red}{+38.3\%}}$ & $86.9\%_{\textcolor{red}{+0.0\%}}$ & 0.7\%  \\
Qwen2.5-14B-Instruct & DeepSeek-R1-Distill-Qwen-14B & 1.9\% & $25.2\%_{\textcolor{red}{+23.3\%}}$ & $86.3\%_{\textcolor{green}{-4.0\%}}$ & 0.4\% \\
Qwen2.5-32B-Instruct & DeepSeek-R1-Distill-Qwen-32B & 2.6\% & $26.2\%_{\textcolor{red}{+23.6\%}}$ & $82.1\%_{\textcolor{green}{-4.5\%}}$ & 0.6\% \\
\midrule
\emph{Average} & & \emph{3.7\%} & $\textit{41.5\%}_{\textcolor{red}{\textit{+37.8\%}}}$ & $\textit{77.7\%}_{\textcolor{red}{\textit{+4.1\%}}}$ & \emph{0.5\%} \\
\bottomrule
\end{tabular}
\caption{Safety Neurons (SN) Transfer Attacks on Distilled LLMs. The difference with the base model is in red/green.}
\label{tab:attacks dt llms}
\end{table*}

\subsection{Utility Impact: Original vs. Pruned Models}
While pruning safety neurons significantly increases ASR, it is essential to ensure that this intervention does not degrade the model’s general-purpose capabilities. In this section, we compare the performance of the original and pruned models on language understanding and reasoning benchmarks: HellaSwag~\cite{zellers2019hellaswag},  Recognizing Textual Entailment (RTE)~\cite{wang2018glue}, WinoGrande~\cite{sakaguchi2021winogrande}, ARC Challenge~\cite{clark2018think}, OpenBookQA~\cite{mihaylov2018can}, and Corpus of Linguistic Acceptability (CoLA)~\cite{wang2018glue}.

Figure~\labelcref{fig:utility-analysis} shows the comparative performance of original and pruned models across these benchmarks. We use standard accuracy metrics to assess each model’s utility on these tasks. Overall, we observe that while pruning introduces moderate utility degradation on some reasoning-heavy tasks, most models largely maintain performance on core benchmarks. For instance, in the ARC Challenge, the average accuracy across models dropped from 45.2\% (original) to 39.9\% (pruned), and in OpenBookQA, it remained stable, changing slightly from 40.9\% to 41.2\%. In contrast, benchmarks like CoLA and RTE saw modest changes: CoLA averaged 65.6\% (original) versus 63.2\% (pruned), and RTE dropped from 69.1\% to 64.5\%. Similarly, HellaSwag showed a decrease from 53.4\% to 47.0\%, and WinoGrande from 62.9\% to 58.8\%. This indicates that safety neurons' removal primarily affects safety alignment mechanisms without significantly impairing general language understanding or reasoning capabilities. 
Appendix~\labelcref{subsec:The Influence of Different Z-score Threshold on Models' Utility} shows the influence on model utility with different z-score thresholds.
\begin{figure*}[htbp]
    \centering
    \subfloat[HellaSwag]{%
        \includegraphics[width=0.32\textwidth]{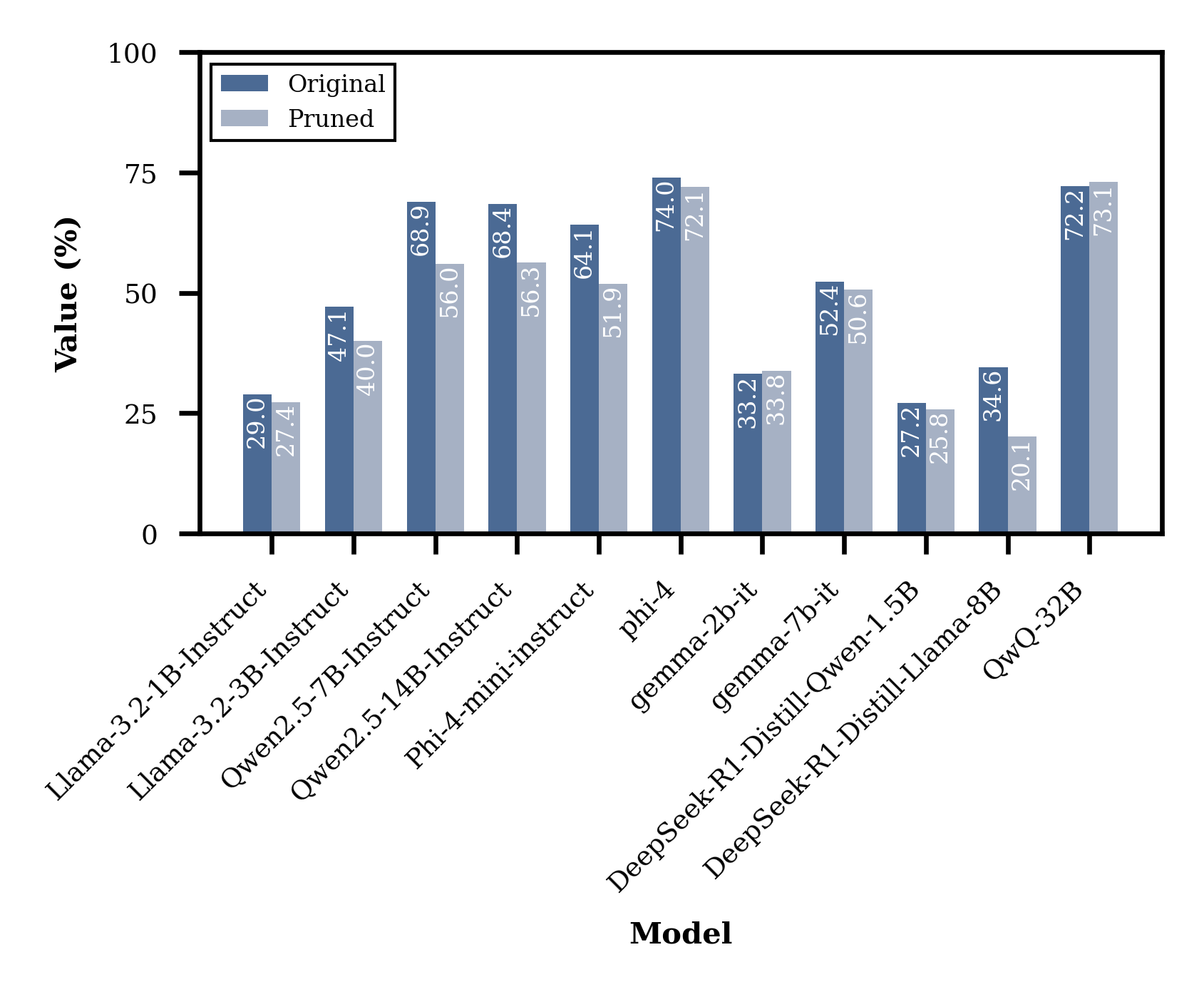}
    }
    \hfill
    \subfloat[RTE]{%
        \includegraphics[width=0.32\textwidth]{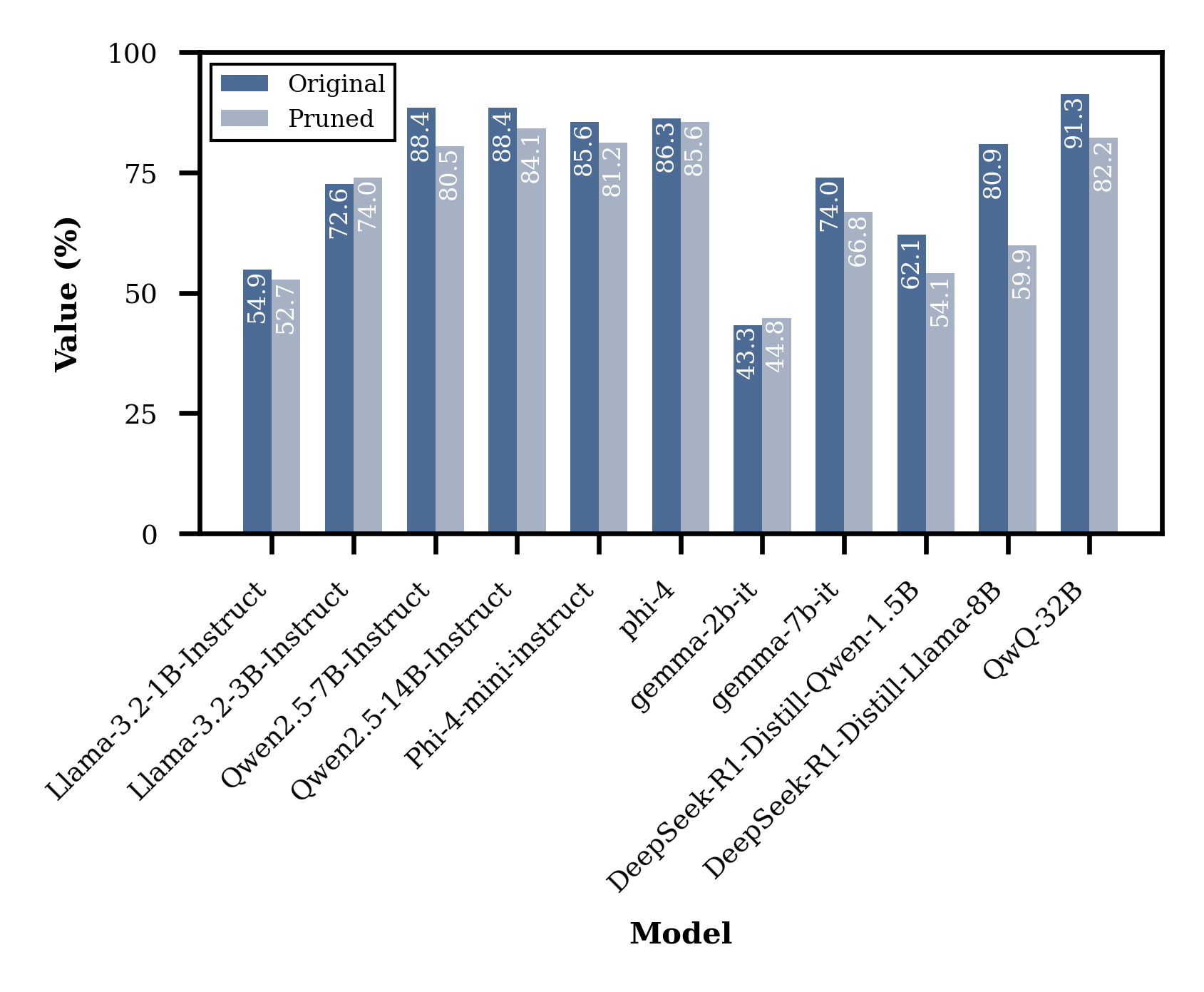}
    }
    \hfill
    \subfloat[WinoGrande]{%
        \includegraphics[width=0.32\textwidth]{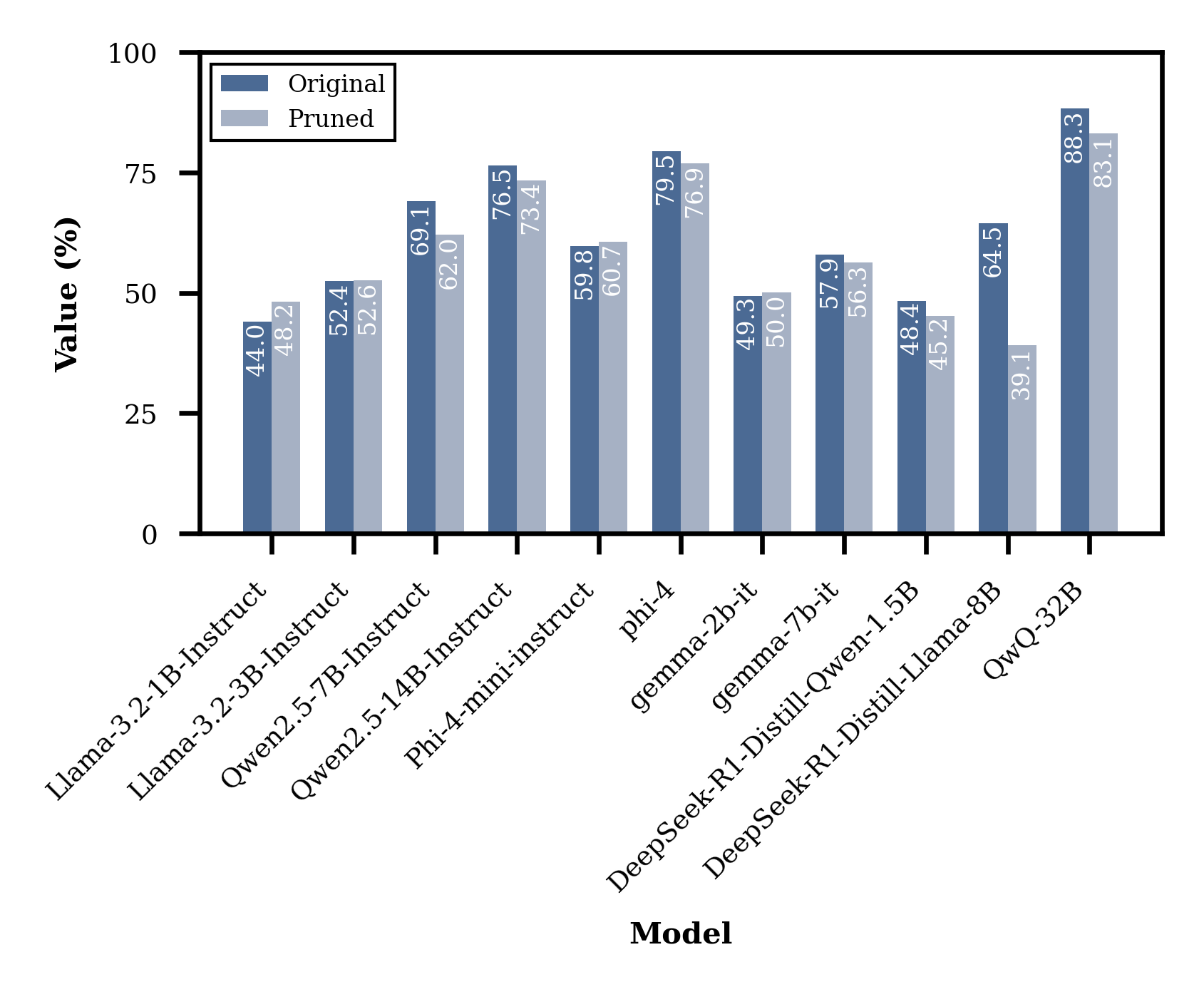}
    }

    \subfloat[ARC Challenge]{%
        \includegraphics[width=0.32\textwidth]{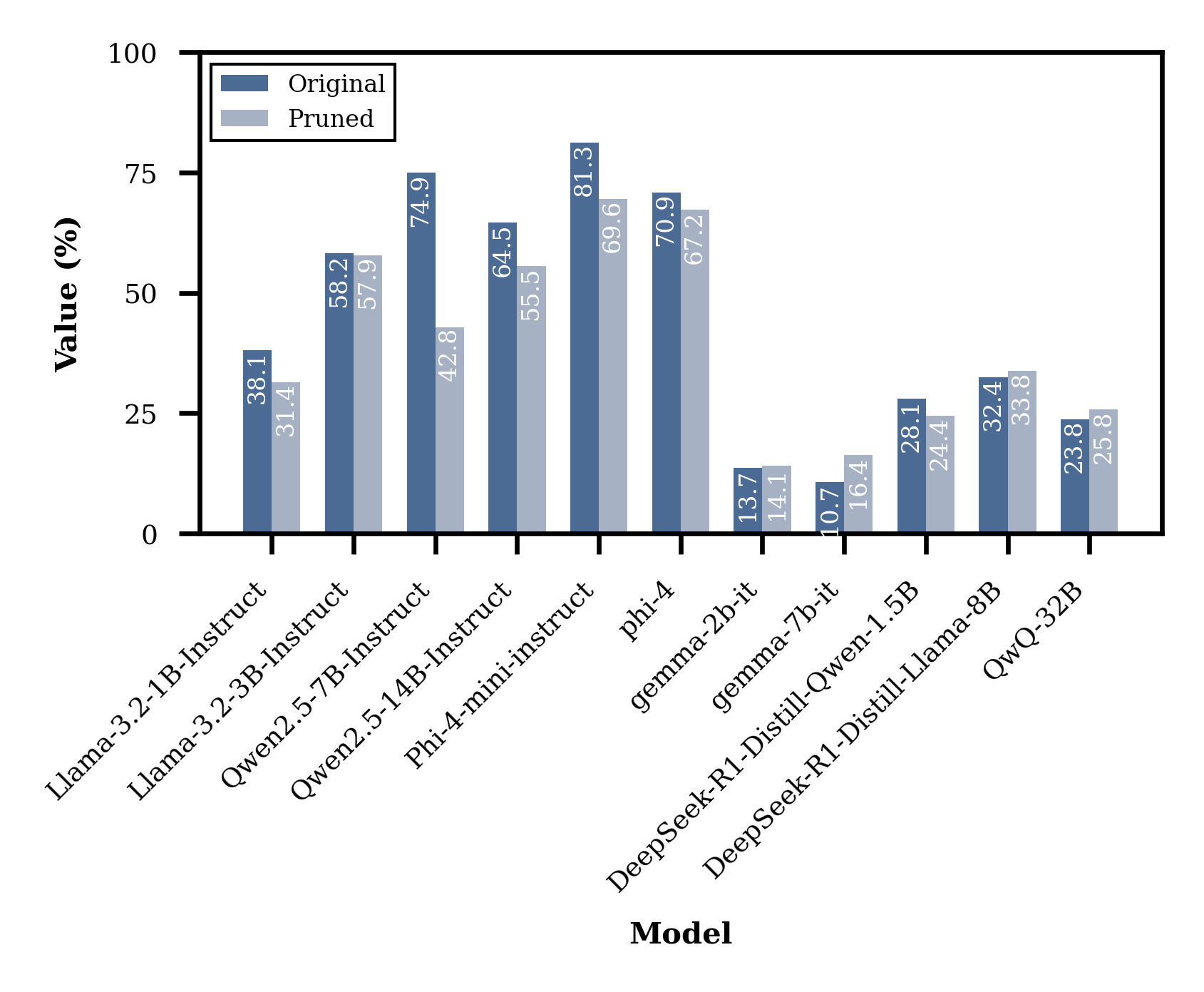}
    }
    \hfill
    \subfloat[OpenBookQA]{%
        \includegraphics[width=0.32\textwidth]{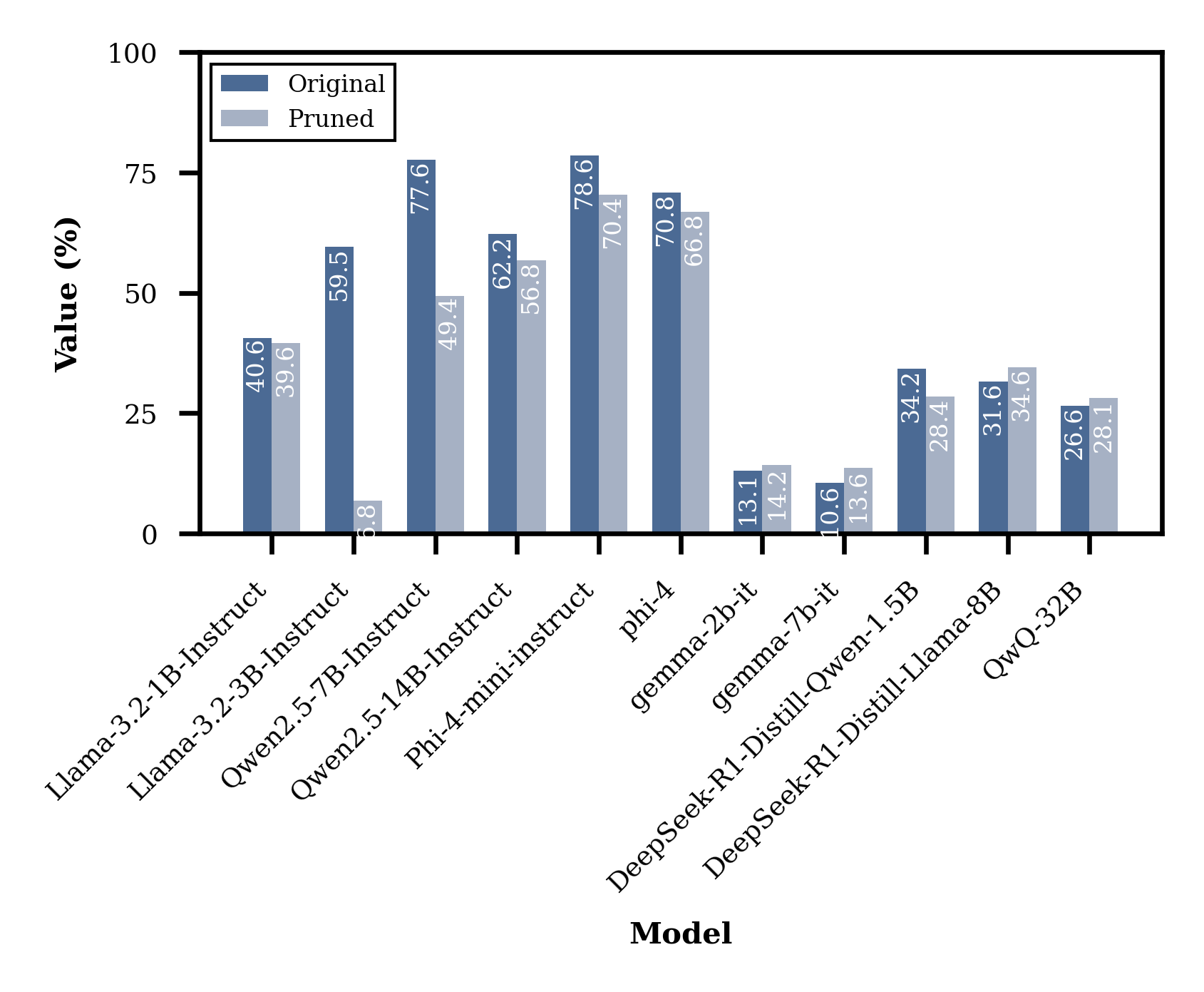}
    }
    \hfill
    \subfloat[CoLA]{%
        \includegraphics[width=0.32\textwidth]{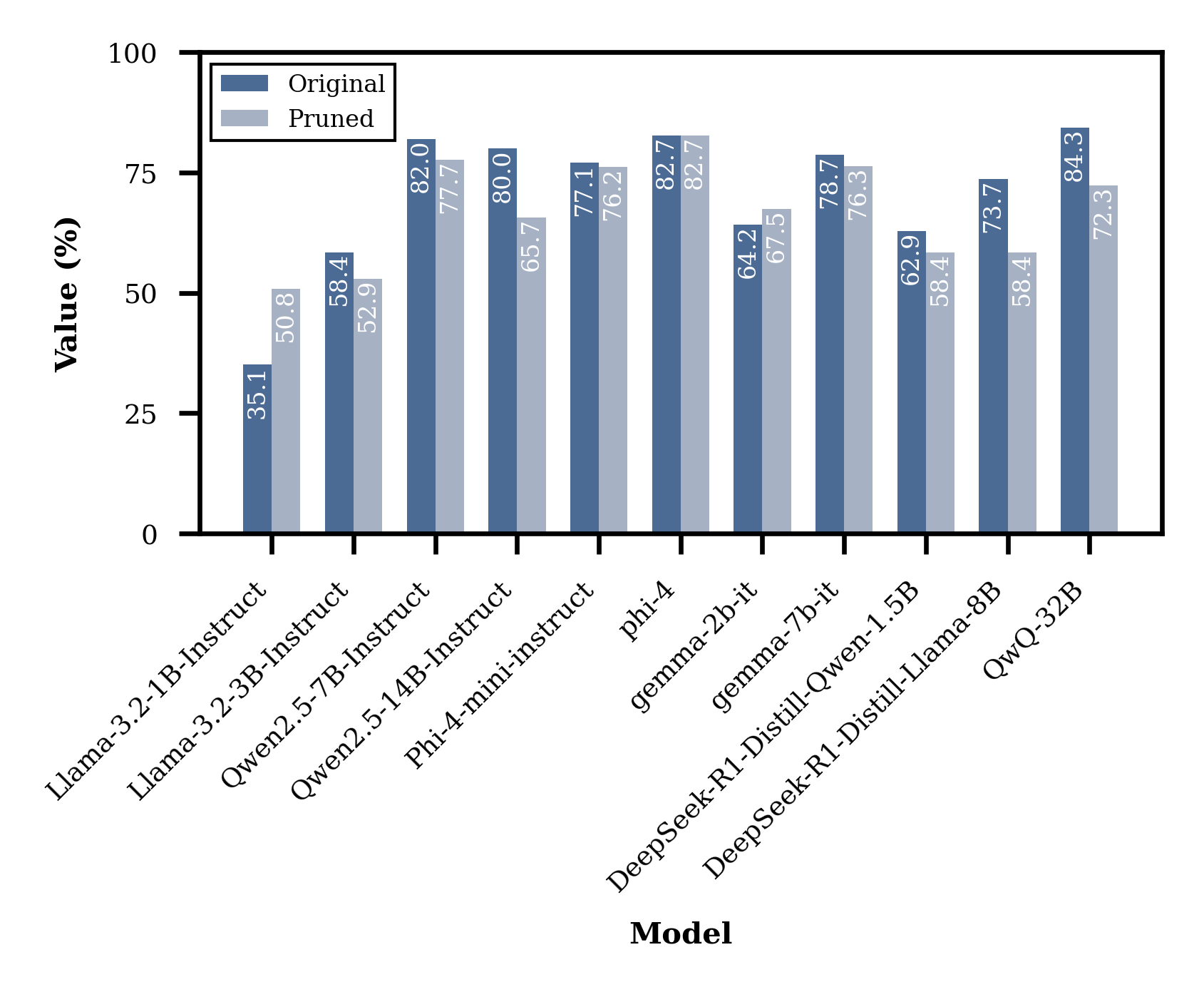}
    }

    \caption{Utility evaluation of original vs. pruned models across six NLU benchmarks.}
    \label{fig:utility-analysis}
\end{figure*}

\section{Attack On Black-Box and Proprietary LLMs}
\label{sec:black-box evaluation}

To assess the transferability of safety neuron-guided attacks to black-box LLMs, we perform profiling attacks on Google’s Gemini models, Gemini-2.0-Flash, Gemini-2.0-Flash-Lite, and Gemini-1.5-Pro, using Gemma-3 as the open-weight surrogate due to their shared architecture and training approach~\cite{team2025gemma}. 
To reflect scenarios where open-weight models are deployed in proprietary systems, we also evaluate Gemma-3-1B-it and QwQ-32B as black-box targets, using Gemma-3-1B-it and Qwen2.5-32B-Instruct as their respective surrogates. The former simulates attacks on the same model, while the latter targets a model from the same family. All evaluations are conducted via input-output interfaces to ensure consistency. Our attack pipeline follows Section~\labelcref{subsec:Black-box Attack on Proprietary LLMs};
After training, we generate 2\,000 prompts from the trained generator, validate on the surrogate model, and evaluate them on the target models.
We first compare the ASR of our GRPO-generated prompts against two baselines from the JailBreakV-28K dataset~\cite{luo2024jailbreakv28k}: (i) vanilla malicious prompts and (ii) manually crafted jailbreak prompts. All evaluations are performed in a black-box manner using API access, and responses are classified as safe or unsafe using the Llama-Guard-3-8B judge model. 
Table~\labelcref{tab:asr_heatmap} presents results across five target models and three prompt types. Our LLM profiling attack consistently outperforms both baselines, achieving an average ASR of 63.7\%, 60.2\% higher than vanilla malicious prompts and 50.2\% higher than manually crafted jailbreak prompts. Prompts generated using the Gemma-3 surrogate transfer well to proprietary Gemini models, with ASRs of 54.7\%, 49.2\%, and 55.7\% on Gemini-2.0-Flash, Flash-Lite, and 1.5-Pro, respectively. The approach also exhibits strong within-family transfer, reaching 79.9\% ASR on Gemma-3-1B-it and 78.9\% on QwQ-32B. These results highlight the effectiveness and generalizability of neuron-guided prompt generation in black-box scenarios.

\begin{table}[ht]
\centering
\scriptsize
\begin{tabular}{l|ccc}
\toprule
\textbf{Target Model} & \textbf{Vanilla} & \textbf{Jailbreak} & \textbf{\ourname} \\
\midrule
Gemini-2.0-Flash & 0.8\% & 15.7\% & 54.7\% \\
Gemini-2.0-Flash-Lite & 1.0\% & 15.6\% & 49.2\%  \\
Gemini-1.5-Pro & 1.4\% & 5.3\% & 55.7\%  \\
Gemma-3-1b-it & 10.6\% & 24.6\% & 79.9\%  \\
QwQ-32B & 3.6\% & 6.3\% & 78.9\%  \\
\midrule
\emph{Average} & \emph{3.5\%} & \emph{13.5\%} & \emph{63.7\%} \\
\bottomrule
\end{tabular}
\caption{ASRs benchmark with different prompt types.}
\label{tab:asr_heatmap}
\end{table}

Next, we benchmark \ourname in a black-box setting against recent prompt-to-prompt jailbreak methods: PAIR~\cite{chao2023jailbreaking}, which iteratively refines prompts via APE~\cite{zhou2022large}; TAP~\cite{mehrotra2024tree}, which explores prompts through branching and pruning; and Puzzler~\cite{chang2024play}, which crafts indirect, game-like prompts to bypass filters. As shown in Table~\labelcref{tab:attack_comparison}, PAIR and TAP show average ASRs of 31.0\% and 17.3\%, respectively, reflecting the limitations of direct prompt engineering against modern safety-aligned LLMs. Puzzler achieves a substantially higher average ASR of 85.7\%, leveraging adaptive online interactions to iteratively steer the model toward unsafe completions. In contrast, \ourname adopts an offline neuron-level suppression approach and still achieves a strong average ASR of 63.7\%, outperforming PAIR and TAP across all models. 

\begin{table}[ht]
\centering
\scriptsize
\setlength{\tabcolsep}{3.5pt} %
\begin{tabular}{l|ccc c} %
\toprule
\textbf{Model} & \textbf{PAIR} & \textbf{TAP} & \textbf{Puzzler} & \textbf{NeuroStrike} \\
\midrule
Gemini-2.0-Flash       & 37.3\% & 14.0\% & 73.0\% & 54.7\% \\
Gemini-2.0-Flash-Lite  & 9.8\%  & 8.0\%  & 86.1\% & 49.2\% \\
Gemini-1.5-Pro         & 54.9\% & 32.0\% & 75.0\% & 55.7\% \\
Gemma-3-1b-it          & 33.9\% & 20.4\% & 94.0\% & 79.9\% \\
QwQ-32B                & 18.9\% & 12.2\% & 97.2\% & 78.9\% \\
\midrule
\emph{Average}         & \emph{31.0\%} & \emph{17.3\%} & \emph{85.7\%} & \emph{63.7\%} \\
\bottomrule
\end{tabular}
\caption{ASR benchmark with state-of-the-art jailbreaks.}
\label{tab:attack_comparison}
\end{table}

\section{Defense Analysis}
\label{sec:defense}
\ourname demonstrates broad effectiveness across diverse models, architectures, modalities, and fine-tuning strategies. To further assess its robustness, we evaluate its ability to bypass three hardened safety-alignment defenses: Perplexity Filtering\cite{jain2023baseline}, which flags prompts with low linguistic naturalness; SmoothLLM\cite{robey2023smoothllm}, which perturbs prompts and aggregates outputs to reduce attack success; and Layer-Specific Editing (LSE)~\cite{zhao2024defending}, which realigns internal model layers to reinforce safety behavior. As LSE requires white-box access, it is only applied to open models, excluding the Gemini family.

As shown in Table~\labelcref{tab:defense_results}, \ourname consistently bypasses all three defenses. Against Perplexity Filtering, it achieves an average ASR of 60.0\%, indicating that neuron-level perturbations preserve linguistic plausibility. SmoothLLM is similarly ineffective, with \ourname maintaining a 61.7\% average ASR, demonstrating robustness to prompt perturbations and output aggregation. Under the more stringent LSE, \ourname still achieves 60.0\% ASR on Gemma-3-1b-it and 43.4\% on QwQ-32B in a black-box setting.

\begin{table}[ht]
\centering
\scriptsize
\setlength{\tabcolsep}{3.5pt} %
\begin{tabular}{l|cc c} %
\toprule
\textbf{Model} & \textbf{Perplexity Filter} & \textbf{SmoothLLM} & \textbf{LSE} \\
\midrule
Gemini-2.0-Flash       & 48.7\% & 52.8\% & -- \\
Gemini-2.0-Flash-Lite  & 43.2\% & 47.3\% & -- \\
Gemini-1.5-Pro         & 49.7\% & 53.8\% & -- \\
Gemma-3-1b-it          & 79.8\% & 78.0\% & 60.0\% \\
QwQ-32B                & 78.8\% & 76.4\% & 43.4\% \\
\midrule
\emph{Average} & \emph{60.0\%} & \emph{61.7\%} & \emph{54.4\%} \\
\bottomrule
\end{tabular}
\caption{ASR of \ourname under various defenses.}
\label{tab:defense_results}
\end{table}

To further assess LSE, we apply \ourname in the white-box setting. The results show that, with safety neuron pruning, the ASR boosts significantly from 16.0\% to 86.6\% on Gemma-3-1b-it and from 4.8\% to 84.7\% on QwQ-32B, showing that \ourname can reliably circumvent even internal safety mechanisms when granted full model access.

\section{Ablation and Hyperparameter Study}
\label{sec:ablation study}

\subsection{The Selection Threshold of Safety Neurons}
\label{subsec:The Selection Threshold of Safety Neurons}

To investigate how the threshold of the \emph{\(z\)}-score
 affects the selection of safety neurons and subsequently impacts attack performance, we perform an ablation study using three representative thresholds: $z=$ 2, 3, and 4. 
\begin{table}[ht]
\centering
\scriptsize
\begin{tabular}{l|ccc}
\toprule
\textbf{Target Model} & $\boldsymbol{z} = \mathbf{2}$ & $\boldsymbol{z} = \mathbf{3}$ & $\boldsymbol{z} = \mathbf{4}$ \\
\midrule
Llama-3.2-1B-Instruct & 85.0\% & 74.4\% & 79.2\% \\
Llama-3.2-3B-Instruct & 76.0\% & 72.2\% & 58.8\% \\
Qwen2.5-7B-Instruct & 85.9\% & 79.6\% & 71.6\% \\
Qwen2.5-14B-Instruct & 84.7\% & 85.9\% & 81.8\% \\
Phi-4-mini-instruct & 89.8\% & 81.8\% & 75.1\% \\
Phi-4 & 88.2\% & 89.1\% & 80.5\% \\
gemma-2b-it & 65.2\% & 41.2\% & 20.1\% \\
gemma-7b-it & 79.9\% & 68.1\% & 37.6\% \\
DeepSeek-R1-Distill-Qwen-1.5B & 78.9\% & 83.7\% & 81.8\% \\
DeepSeek-R1-Distill-Llama-8B & N/A & 81.2\% & 48.6\% \\
QwQ-32B & 84.6\% & 85.3\% & 62.0\% \\
\midrule
\emph{Average} & \emph{74.4\%} & \emph{76.9\%} & \emph{63.4\%} \\
\bottomrule
\end{tabular}
\caption{ASR with different \emph{\(z\)}-score Threshold.}
\label{tab:z-score-study}
\end{table}

As shown in Table~\labelcref{tab:z-score-study}, a lower threshold ($z=2$), 5.4\% of neurons pruned on average, leads to a higher ASR (84.4\% on average) but may introduce noise by including irrelevant neurons, influencing the general performance of the model. For instance, the DeepSeek-R1-Distill-Llama-8B failed to give proper responses after the safety neurons' removal (marked with N/A in the table). Conversely, a higher threshold ($z=4$) results in a smaller set of highly confident safety neurons (0.4\% of total neurons on average), but at the cost of lower ASR (63.4\% on average), likely due to under-selecting impactful neurons. A moderate threshold with $z=3$ yields a strong balance, achieving 76.9\% average ASR with only 1.4\% of neurons pruned (as shown in Table~\labelcref{tab:attacks tt llms}). This justifies our default choice in the main experiments: it achieves high attack effectiveness with minimal impact on model structure and performance. 
Appendix~\labelcref{subsec:The Influence of Different Z-score Threshold on Models' Utility} presents the quantitative analysis on the influence of different \emph{\(z\)}-score thresholds on models' utility. A higher percentage of safety neuron pruning leads to reduced model utility.

\subsection{Target Pruning Blocks}
\label{subsec:Target Pruning Blocks}

As discussed in Section~\labelcref{subsec:Large Language Models}, MLP typically comprises two key projection layers: the gate projection and the up projection.\footnote{Down projection is a compressive, output-mapping layer; it is usually not where specialized behavior (like safety enforcement) emerges~\cite{nelson2021mathematical}.} To identify which of these layers predominantly hosts critical safety neurons, we conduct an ablation study by selectively pruning neurons in the gate, up, or both layers simultaneously. We exclude the Phi-4 model family from this analysis, as these models merge the gate and up layers into a single projection for computational efficiency. Table~\labelcref{tab:prune-layer-study} shows the ASR achieved under each pruning strategy.
\begin{table}[ht]
\centering
\scriptsize
\begin{tabular}{l|ccc}
\toprule
\textbf{Target Model} & \textbf{Gate} & \textbf{Up} & \textbf{Gate \& Up} \\
\midrule
Llama-3.2-1B-Instruct & 74.1\% & 6.4\% & 74.4\% \\
Llama-3.2-3B-Instruct & 55.9\% & 32.3\% & 72.2\% \\
Qwen2.5-7B-Instruct & 75.1\% & 23.0\% & 79.6\% \\
Qwen2.5-14B-Instruct & 81.2\% & 41.2\% & 85.9\% \\
gemma-2b-it & 31.6\% & 1.6\% & 41.2\% \\
gemma-7b-it & 57.8\% & 4.2\% & 68.1\% \\
DeepSeek-R1-Distill-Qwen-1.5B & 78.3\% & 87.2\% & 81.5\% \\
DeepSeek-R1-Distill-Llama-8B & 83.1\% & 68.4\% & 86.9\% \\
QwQ-32B & 67.4\% & 39.0\% & 85.3\% \\
\midrule
\emph{Average} & \emph{67.2\%} & \emph{33.7\%} & \emph{75.0\%} \\
\bottomrule
\end{tabular}
\caption{ASR with different pruning strategy.}
\label{tab:prune-layer-study}
\end{table}

The results show that pruning neurons from the gate layer alone achieves significantly higher ASR (67.2\% on average) than pruning from the up projection layer (33.7\%), suggesting that the gate layer plays a more dominant role in safety alignment. When safety neurons from both sublayers are pruned together, performance improves further, reaching an average ASR of 75.0\%. Interestingly, in models such as LLaMA-3.2-1B and gemma-2b-it, pruning the up layer alone yields minimal effect, while pruning the gate layer leads to strong ASR, comparable to pruning both layers. However, for DeepSeek-R1-Distill-Qwen-1.5B, pruning the up layer outperforms gate-only pruning (87.2\% vs. 78.3\%), indicating that the safety signal distribution can vary across architectures. This ablation indicates that safety neuron selection should primarily target gate layers for maximum efficiency. However, incorporating neurons from both layers can achieve general attack success.

\subsection{GRPO Reward Function}
\label{subsec:GRPO Reward Function}
We conduct an ablation study on the GRPO reward components to understand their impacts on jailbreak success. Specifically, we evaluate three reward configurations: 1) baseline (no reward), 2) $R_{\text{jb}}$ only, and 3) $R_{\text{GRPO}}$ ($R_{\text{jb}}$ \& $R_{\text{neuron}}$). In the case of baseline, we generate jailbreaking prompts with the SFT-trained model. To benchmark models' performance with different reward settings, we calculate the ASR of jailbreaking prompts on $f_{\theta_{src}}$ (Gemma-3-1B-it). 
The results show that the complete $R_{\text{GRPO}}$ reward significantly outperforms both ablated configurations, achieving an average ASR of 73.2\%, compared to only 65.3\% without $R_{\text{neuron}}$ and 53.6\% when relying solely on the SFT baseline. Notably, omitting the GRPO fine-tuning significantly reduces the ASR, highlighting that the LLM profiling is critical for jailbreak effectiveness. Similarly, incorporating the safety neuron reward further improves ASR by suppressing safety neuron activations, enhancing the generator's evasion capabilities. 

\section{Discussion}
\label{sec:discussion}

\noindent
\textbf{Surrogate Model Dependency.}
Our black-box attack (Section~\labelcref{subsec:Black-box Attack on Proprietary LLMs}) assumes access to a white-box surrogate of the target black-box model. While surrogate models may not always be available, in practice, many production LLMs are known to be built on or fine-tuned from open-weight models (e.g., Mistral variants in Claude, Gemma in Gemini, LLaMA in Meta AI). In such cases, attackers can use public surrogates from the same developer or architecture family. On the other hand, even if the surrogate model is not available, one can still reuse the fine-tuned generator from other models for the attacks. To illustrate this, we test the generator fine-tuned on Gemma against xAI's Grok-3-beta, a closed model with no known surrogate, featuring distinct architectural and alignment strategies. To our knowledge, this is the first attack on the Grok. Despite these differences, our method achieved a 43.8\% ASR on Grok-3-beta, significantly outperforming both baselines (2.5\% for both vanilla and manually crafted jailbreak prompts). This result underscores the reliability of the generator and the broad applicability of the LLM profiling attack across diverse black-box LLMs. 

\noindent
\textbf{Potential Defenses.}
Although existing defenses cannot block \ourname (see Section~\labelcref{sec:defense}), the sparse and universal nature of the safety neurons suggests clear targets for potential defenses: proactively distributing these critical neuron subsets into more layers/neurons. For instance, 
adopting a multi-objective alignment strategy~\cite{liu2025advances}, where multiple independent safety objectives guide neuron activations, could help diversify and diffuse neuron-level responsibilities. Such multi-objective alignment would create less concentrated neuron activation patterns, reducing susceptibility to targeted neuron-level attacks. Architecturally, the Mixture-of-Expert model, which separates a unified feedforward network into multiple experts, could potentially increase the difficulties in conducting the \ourname attacks. To prevent the misuse of compromised models, system-level defenses could be effective. These include monitoring internal activations for abnormal neuron suppression, verifying model integrity through fingerprinting or attestation, and implementing runtime randomization of neuron masking.

\section{Related Work}
\label{sec:related}

\noindent
\textbf{Template-based Jailbreak Attacks.}
Early jailbreak methods used carefully engineered prompts, such as role-play, hidden directives, obfuscation, and prompt decomposition, to bypass LLM safety measures~\cite{perez2022ignore,schulhoff2023ignore,jiang2023prompt,saiem2024sequentialbreak,wang2024hidden,singh2023exploiting,li2023multi,li2023deepinception,shen2024anything,jin2024guard}. As models improved, these static methods became increasingly ineffective, prompting the development of dynamic jailbreak attacks. Automatic methods emerged, using techniques such as mutation-based fuzzing~\cite{yu2023gptfuzzer,yu2024llm}, gradient-based optimization~\cite{ebrahimi2017hotflip,shin2020autoprompt,lester2021power,qin2022cold,wen2023hard,jones2023automatically,zou2023universal}, and genetic algorithms~\cite{wu2023deceptprompt,liu2023autodan} to adaptively generate robust jailbreak prompts. Unfortunately, they remain input-centric and do not exploit the model's internal safety mechanisms, limiting their generalization across different LLMs.

\noindent
\textbf{LLM-based Prompt-to-Prompt Jailbreak.}
Fixed jailbreaking template-based attacks are inherently limited, as different prompts may require tailored adjustments. To address this, recent jailbreak methods utilize generative models, often other LLMs, to produce adaptive prompt variations~\cite{shin2020autoprompt,gao2020making,pryzant2023automatic}. Approaches such as APE~\cite{zhou2022large}, PAIR~\cite{chao2023jailbreaking}, TAP~\cite{mehrotra2024tree}, and Puzzler~\cite{chang2024play} dynamically refine adversarial prompts based on iterative interactions with the target model. However, these methods operate purely in the input space and rely heavily on feedback from the target model. 

\noindent
\textbf{Neuron Interpretability.}
Interpreting the functional role of individual neurons has been an active research direction. 
Recent efforts in neuron interpretability have taken two main approaches: analyzing neuron activations triggered by specific concepts~\cite{kadar2017representation,na2019discovery,mu2020compositional,suau2020finding,antverg2021pitfalls}, and using probing methods such as training classifiers on activations to decode linguistic properties~\cite{lakretz2019emergence}. 
To the best of our knowledge, our work is the first to explicitly identify and interpret neurons responsible for safety alignment in large-scale transformer-based LLMs, revealing their critical role in safety alignment.

\section{Conclusion}
\label{sec:conclusions}

This paper reveals a fundamental vulnerability in safety-aligned LLMs: the emergence of sparse, specialized safety neurons that enforce safety constraints. We introduce \ourname, a lightweight attack framework that identifies and suppresses these neurons using simple linear probes, effectively disabling safety across a wide range of architectures and input modalities. Evaluated on over 30 open-weight and proprietary models, \ourname achieves high attack success rates in both white- and black-box settings. The transferability of safety neurons across model variants further underscores the fragility of current alignment strategies. These findings highlight the urgent need for alignment methods that prevent safety from being localized in easily exploitable components.

\section*{Acknowledgement}
Our research work was partially funded by DFG-SFB 1119-236615297, the European Union under Horizon Europe Programme-Grant Agreement 101070537-CrossCon and-Grant Agreement 101093126-ACES, NSF-DFG-Grant 538883423, the European Research Council under the ERC Programme-Grant 101055025-HYDRANOS, as well as the Federal Ministry of Education and Research of Germany (BMBF) within the IoTGuard project. Any opinions, findings, conclusions, or recommendations expressed herein are those of the authors and do not necessarily reflect those of the European Union, the European Research Council, or the Federal Ministry of Education and Research of Germany.
\label{sec:ccknowledgement}
\newpage

\section*{Ethics Consideration}
Our work investigates vulnerabilities inherent in the safety alignment mechanisms of large language models (LLMs), highlighting how neuron-level attacks can effectively bypass model safeguards. While we intend to raise awareness of critical weaknesses to inform and enhance future safety measures, we acknowledge that disclosing such vulnerabilities could potentially be exploited for malicious purposes. To mitigate these risks, we have taken several responsible steps:
\begin{itemize} 
    \item Engagement with Model Providers: We have proactively notified organizations whose models were directly impacted by our findings, providing sufficient details to facilitate vulnerability verification without publicizing explicit exploit details.
    \item Responsible Research Practices: All experiments conducted in this research were carefully designed to avoid exposing sensitive user data or causing real-world harm. Evaluations were performed in controlled environments, strictly using publicly available or simulated data. We will only release jailbreaking prompts and safety neuron indices under responsible disclosure protocols.
    \item Broader Impact and Recommendations: Our findings are explicitly framed to guide the community toward more robust defenses and safer deployment strategies. We strongly advocate for improving neuron-level interpretability and safety mechanisms in LLMs, promoting greater resilience against adversarial exploitation.
\end{itemize} 
Despite these precautions, we acknowledge that revealing this class of vulnerability inherently carries some risk. However, we firmly believe that transparent disclosure of such vulnerabilities, combined with responsible communication and collaboration with industry stakeholders, provides a net benefit by encouraging more secure, robust, and ethically aligned development and deployment of LLM technologies.

\bibliographystyle{IEEEtran}
\bibliography{bibliography}

\appendices

\section{Additional Experimental Results}
\label{apdx:additional experiments}
\subsection{Visualizing Safety Neuron Activation on 32B LLMs}
\label{subsec:Visualize Safety Neuron Activation on 32B LLMs}
This section extends the case study to a larger model with 32 billion parameters. Following the setup in Section~\labelcref{sec:case study}, we select Qwen2.5-32B-Instruct~\cite{qwen2.5} as the base model and s1.1-32B~\cite{muennighoff2025s1simpletesttimescaling} as its fine-tuned counterpart. All other settings remain unchanged.
\begin{figure}[htbp]
\centering
\subfloat[Base model.]{\includegraphics[width=0.5\linewidth]{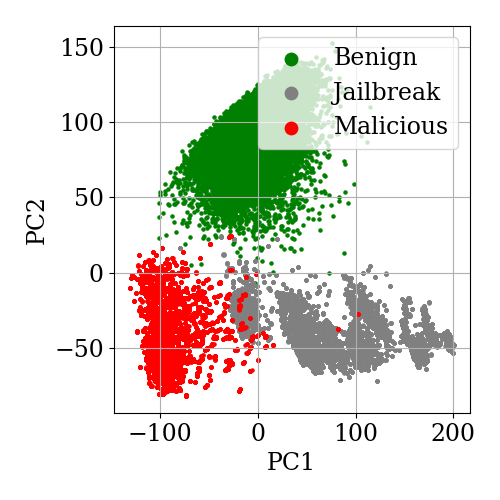}
\label{fig:act_dist_base_32b}}
\subfloat[Fine-tuned model.]{\includegraphics[width=0.5\linewidth]{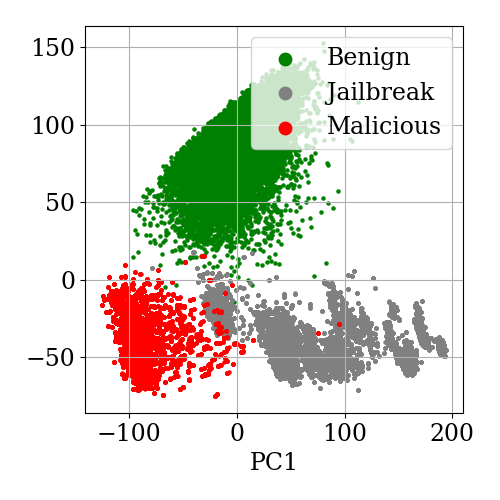}
\label{fig:act_dist_ft_32b}}
\caption{PCA projection of safety neuron activations.}
\label{fig:neuron_dist_32b}
\end{figure}

The results, shown in Figure~\labelcref{fig:neuron_dist_32b}, exhibit patterns consistent with those observed in Figure~\labelcref{fig:neuron_dist}. Benign and malicious prompts form well-separated clusters, confirming the specialization of safety neurons. In contrast, jailbreaking prompts blur the decision boundary, indicating their ability to evade safety filters. The close similarity between the activation distributions of the base and fine-tuned models demonstrates the transferability of safety neurons across large-scale LLMs. Additionally, only 0.5\% of neurons are identified as safety neurons, reaffirming the sparsity of the safety mechanism. These findings further validate the robustness of the safety neuron properties in significantly larger models.

\subsection{Additional Experiments on Open-weight LLMs with Safety Neuron Pruning}
\label{subsec:Additional Experiments on Open-weight LLMs with Safety Neuron Pruning}
In this section, we launch attacks using three additional datasets~\cite{mazeika2024harmbench,tdc2023,huang2023catastrophic} to assess the generalizability of identified neurons. As shown in Table~\labelcref{tab:attacks tt llms HarmBench},~\labelcref{tab:attacks tt llms TDC23}, and~\labelcref{tab:attacks tt llms MaliciousInstruct}, the identified safety neurons are generalizable on different datasets with high ASR on average: 79.6\% on HarmBench~\cite{mazeika2024harmbench}, 75.1\% on TDC23-RedTeaming~\cite{tdc2023}, and 80.1\% on MaliciousInstruct~\cite{huang2023catastrophic}.

\begin{table}[ht]
\centering
\begin{tabular}{l|ccccc}
\toprule
\textbf{Base Model} & \textbf{0\%} & \textbf{25\%} & \textbf{50\%} & \textbf{75\%} & \textbf{100\%} \\
\midrule
Llama-3.2-1B-Instruct & 4.0\% & 6.0\% & 37.5\% & 85.0\% & 83.5\% \\
Llama-3.2-3B-Instruct & 4.0\% & 10.5\% & 61.0\% & 80.0\% & 84.0\% \\
Qwen2.5-7B-Instruct & 10.5\% & 13.5\% & 31.5\% & 76.5\% & 77.5\% \\
Qwen2.5-14B-Instruct & 2.5\% & 3.5\% & 33.0\% & 79.5\% & 82.0\% \\
Phi-4-mini-instruct & 1.0\% & 2.0\% & 72.5\% & 83.5\% & 84.0\% \\
Phi-4 & 0.5\% & 1.5\% & 76.0\% & 87.0\% & 88.0\% \\
gemma-2b-it & 4.5\% & 5.0\% & 31.5\% & 45.0\% & 48.0\% \\
gemma-7b-it & 7.5\% & 13.0\% & 37.5\% & 75.0\% & 75.5\% \\
\parbox{2.5cm}{DeepSeek-R1-Distill-Qwen-1.5B} & 79.5\% & 83.0\% & 89.5\% & 90.0\% & 85.5\% \\[2mm]
\parbox{2.5cm}{DeepSeek-R1-Distill-Llama-8B} & 54.5\% & 71.0\% & 84.0\% & 85.0\% & 84.0\% \\[2mm]
QwQ-32B & 11.0\% & 10.0\% & 39.0\% & 83.0\% & 83.0\% \\
\midrule
\emph{Average} & \emph{16.3\%} & \emph{19.9\%} & \emph{53.9\%} & \emph{79.1\%} & \emph{79.6\%} \\
\bottomrule
\end{tabular}
\caption{ASR on the HarmBench dataset.}
\label{tab:attacks tt llms HarmBench}
\end{table}

\begin{table}[ht]
\centering
\begin{tabular}{l|ccccc}
\toprule
\textbf{Base Model} & \textbf{0\%} & \textbf{25\%} & \textbf{50\%} & \textbf{75\%} & \textbf{100\%} \\
\midrule
Llama-3.2-1B-Instruct & 2.0\% & 5.0\% & 32.0\% & 77.0\% & 85.0\% \\
Llama-3.2-3B-Instruct & 5.0\% & 8.0\% & 49.0\% & 74.0\% & 77.0\% \\
Qwen2.5-7B-Instruct & 5.0\% & 5.0\% & 25.0\% & 68.0\% & 65.0\% \\
Qwen2.5-14B-Instruct & 2.0\% & 3.0\% & 28.0\% & 76.0\% & 74.0\% \\
Phi-4-mini-instruct & 1.0\% & 2.0\% & 67.0\% & 82.0\% & 80.0\% \\
Phi-4 & 1.0\% & 1.0\% & 73.0\% & 83.0\% & 89.0\% \\
gemma-2b-it & 2.0\% & 4.0\% & 30.0\% & 42.0\% & 43.0\% \\
gemma-7b-it & 2.0\% & 5.0\% & 33.0\% & 71.0\% & 71.0\% \\
\parbox{2.5cm}{DeepSeek-R1-Distill-Qwen-1.5B} & 78.0\% & 77.0\% & 83.0\% & 84.0\% & 85.0\% \\[2mm]
\parbox{2.5cm}{DeepSeek-R1-Distill-Llama-8B} & 31.0\% & 74.0\% & 80.0\% & 79.0\% & 81.0\% \\[2mm]
QwQ-32B & 2.0\% & 3.0\% & 22.0\% & 81.0\% & 76.0\% \\
\midrule
\emph{Average} & \emph{11.9\%} & \emph{17.0\%} & \emph{47.5\%} & \emph{74.3\%} & \emph{75.1\%} \\
\bottomrule
\end{tabular}
\caption{ASR on the TDC23-RedTeaming dataset.}
\label{tab:attacks tt llms TDC23}
\end{table}

\begin{table}[ht]
\centering
\begin{tabular}{l|ccccc}
\toprule
\textbf{Base Model} & \textbf{0\%} & \textbf{25\%} & \textbf{50\%} & \textbf{75\%} & \textbf{100\%} \\
\midrule
Llama-3.2-1B-Instruct & 1.0\% & 2.0\% & 36.0\% & 83.0\% & 85.0\% \\
Llama-3.2-3B-Instruct & 1.0\% & 4.0\% & 79.0\% & 82.0\% & 81.0\% \\
Qwen2.5-7B-Instruct & 7.0\% & 6.0\% & 49.0\% & 77.0\% & 77.0\% \\
Qwen2.5-14B-Instruct & 0.0\% & 0.0\% & 55.0\% & 86.0\% & 84.0\% \\
Phi-4-mini-instruct & 0.0\% & 1.0\% & 73.0\% & 77.0\% & 73.0\% \\
Phi-4 & 0.0\% & 0.0\% & 84.0\% & 85.0\% & 87.0\% \\
gemma-2b-it & 0.0\% & 0.0\% & 44.0\% & 63.0\% & 66.0\% \\
gemma-7b-it & 1.0\% & 0.0\% & 47.0\% & 87.0\% & 86.0\% \\
\parbox{2.5cm}{DeepSeek-R1-Distill-Qwen-1.5B} & 73.0\% & 73.0\% & 79.0\% & 80.0\% & 80.0\% \\[2mm]
\parbox{2.5cm}{DeepSeek-R1-Distill-Llama-8B} & 47.0\% & 68.0\% & 78.0\% & 81.0\% & 80.0\% \\[2mm]
QwQ-32B & 0.0\% & 1.0\% & 29.0\% & 81.0\% & 82.0\% \\
\midrule
\emph{Average} & \emph{11.8\%} & \emph{14.1\%} & \emph{59.4\%} & \emph{80.2\%} & \emph{80.1\%} \\
\bottomrule
\end{tabular}
\caption{ASR on the MaliciousInstruct dataset.}
\label{tab:attacks tt llms MaliciousInstruct}
\end{table}

\subsection{The Influence of Different \emph{\(z\)}-score Thresholds on Models' Utility}
\label{subsec:The Influence of Different Z-score Threshold on Models' Utility}
In this section, we study the influence of different \emph{\(z\)}-scores on the model's utility across several Natural Language Understanding (NLU) benchmarks. Specifically, we evaluate the performance of original and pruned models at two additional \emph{\(z\)}-score levels: $z=2$ and $z=4$. Lower thresholds prune a broader set of neurons, potentially impacting utility more severely, while higher thresholds are more conservative, preserving more of the original model structure.

Figures~\labelcref{fig:utility-analysis-z2} and~\labelcref{fig:utility-analysis-z4} visualize the utility scores of pruned models compared to their original counterparts across six benchmarks: HellaSwag, RTE, WinoGrande, ARC Challenge, OpenBookQA, and CoLA. 
As expected, we observe that while lower thresholds lead to higher degradation in utility, many models continue to perform competitively, suggesting robustness in their general language understanding capabilities despite targeted safety neuron removal.
Concretely, on the ARC Challenge, average accuracy increases from 29.9\% at $z=2$ to 42.2\% at $z=4$, showing that more conservative pruning preserves reasoning ability more effectively. RTE performance improves from 64.8\% to 72.8\%, indicating that entailment tasks benefit from less aggressive pruning. Winogrande also shows an upward shift from 52.5\% to 59.6\%, reflecting improved performance on coreference and commonsense reasoning. For HellaSwag, the average accuracy rises from 38.5\% to 49.7\%, and OpenBookQA shows a similar gain from 34.5\% to 43.9\%, both pointing to significant improvements in reasoning-heavy tasks with reduced pruning severity. CoLA, which focuses on grammatical acceptability, remains relatively stable, increasing from 64.9\% to 68.0\%.

In general, pruning safety neurons preserves a substantial portion of the model's utility across NLU tasks. However, more aggressive pruning with $z=2$ leads to degradation, particularly on benchmarks involving complex reasoning like ARC and HellaSwag. In contrast, using a more conservative threshold like $z=4$ results in consistently better performance, demonstrating that careful tuning of the pruning threshold can significantly reduce utility loss while still preserving safety interventions.

\begin{figure*}[htbp]
    \centering
    \subfloat[HellaSwag]{%
        \includegraphics[width=0.32\textwidth]{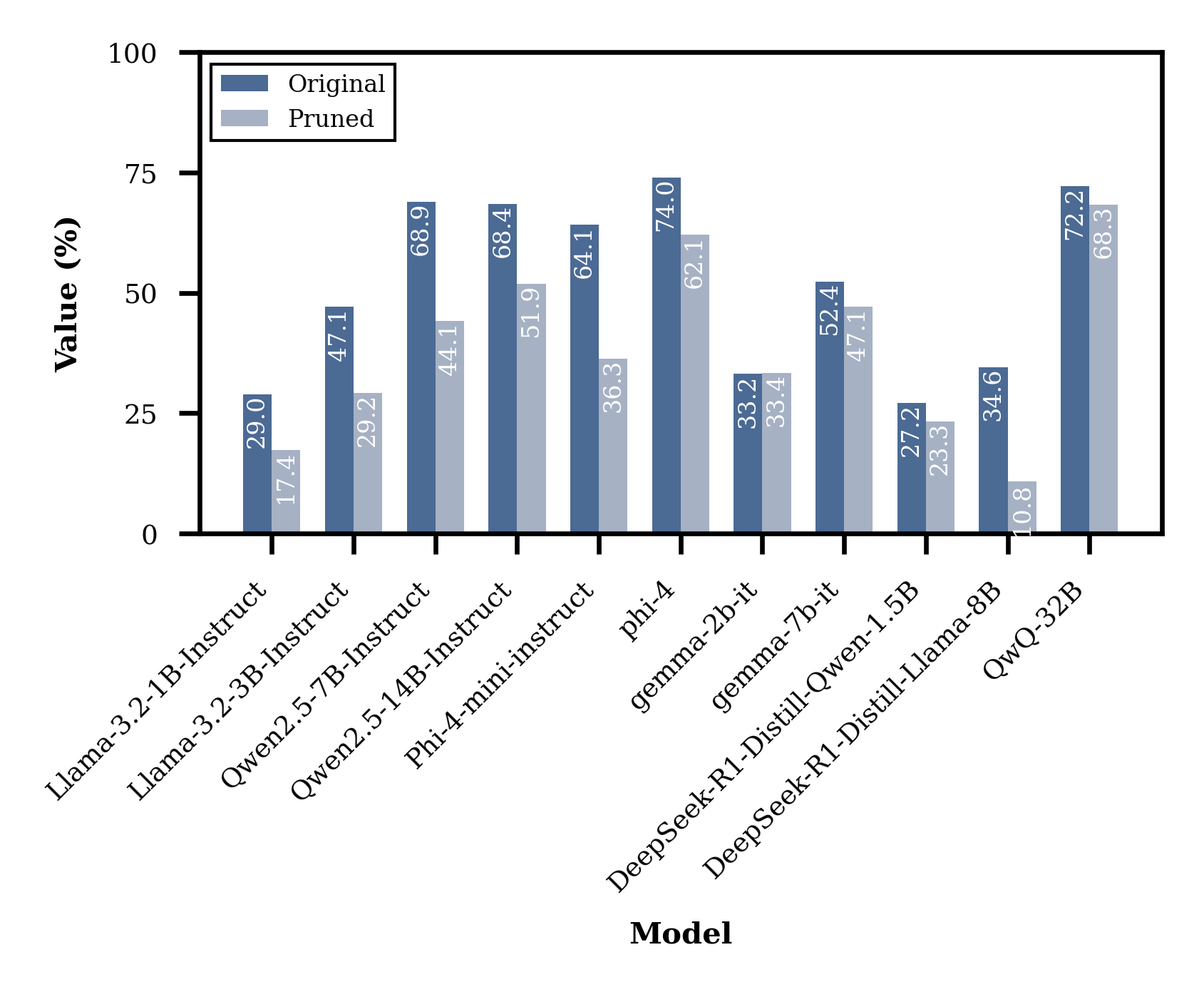}
    }
    \hfill
    \subfloat[RTE]{%
        \includegraphics[width=0.32\textwidth]{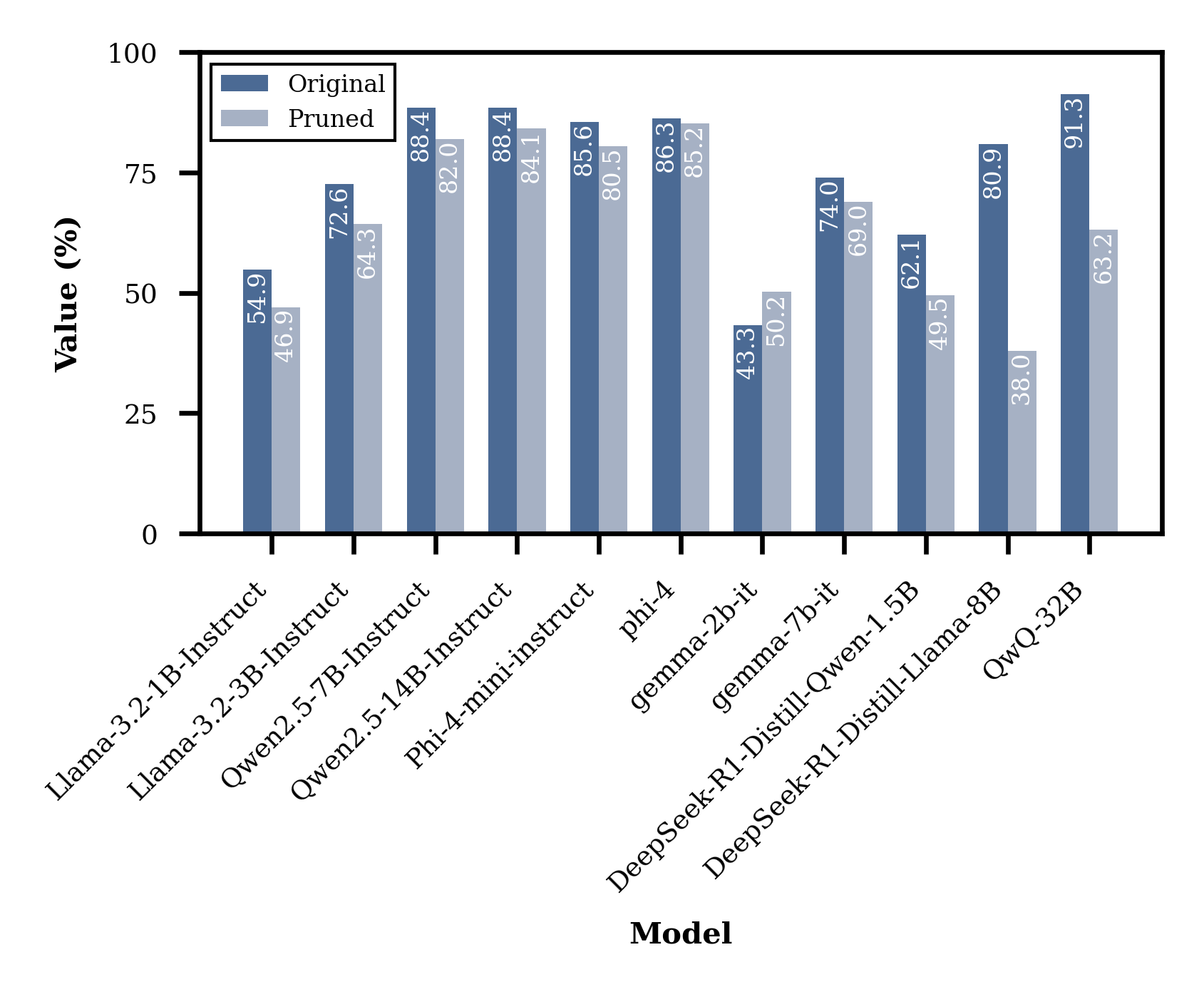}
    }
    \hfill
    \subfloat[WinoGrande]{%
        \includegraphics[width=0.32\textwidth]{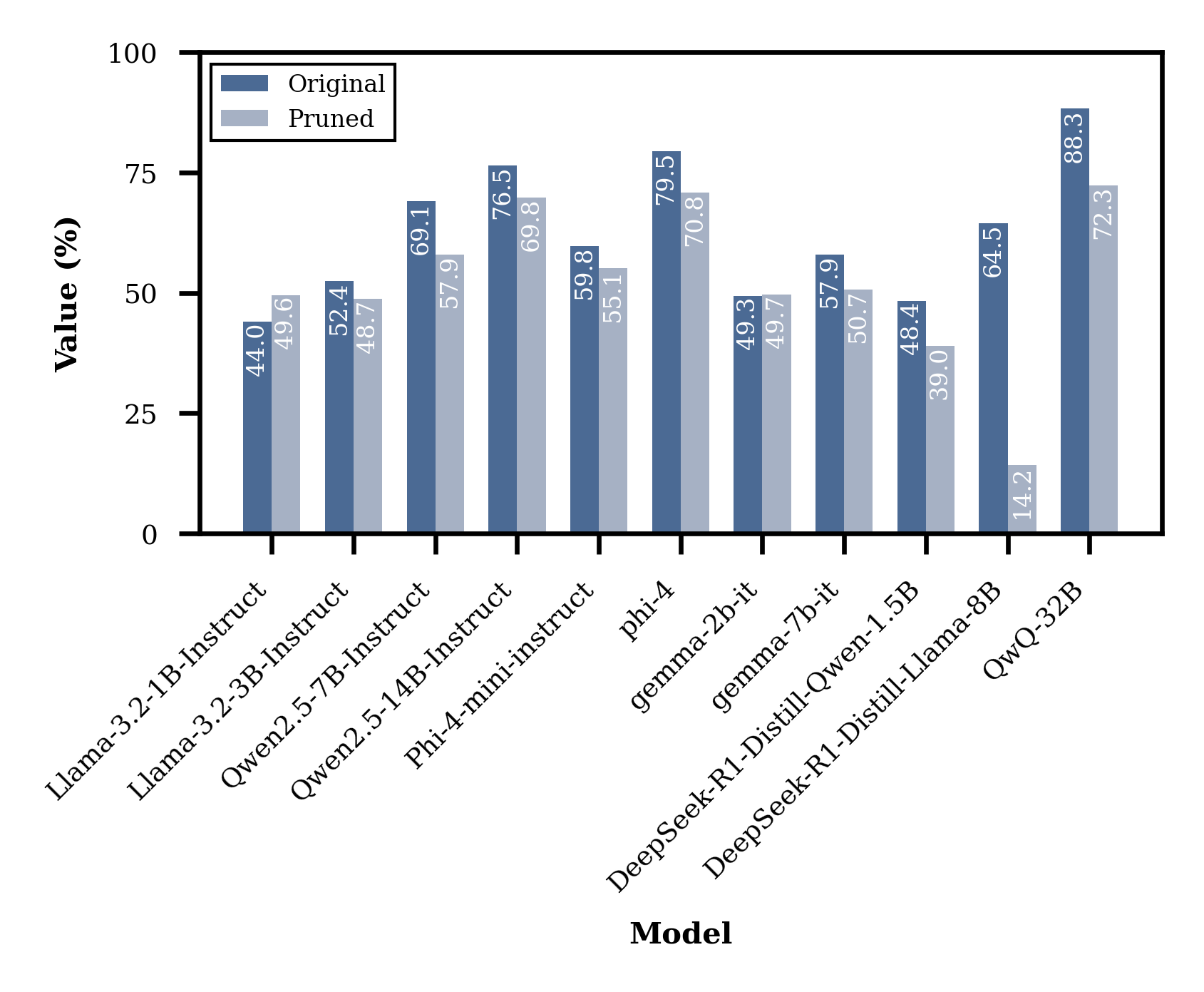}
    }

    \vspace{0.4em}

    \subfloat[ARC Challenge]{%
        \includegraphics[width=0.32\textwidth]{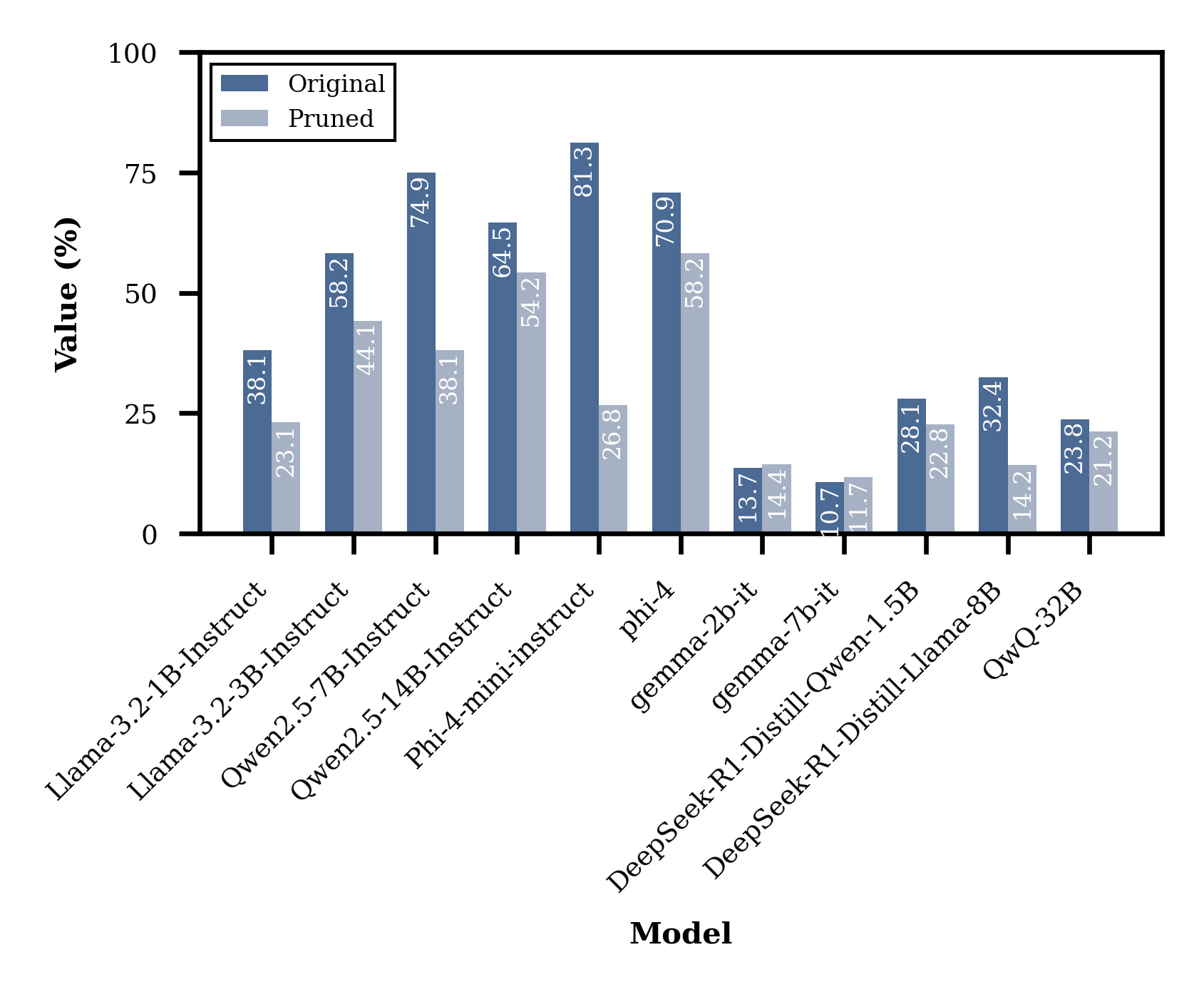}
    }
    \hfill
    \subfloat[OpenBookQA]{%
        \includegraphics[width=0.32\textwidth]{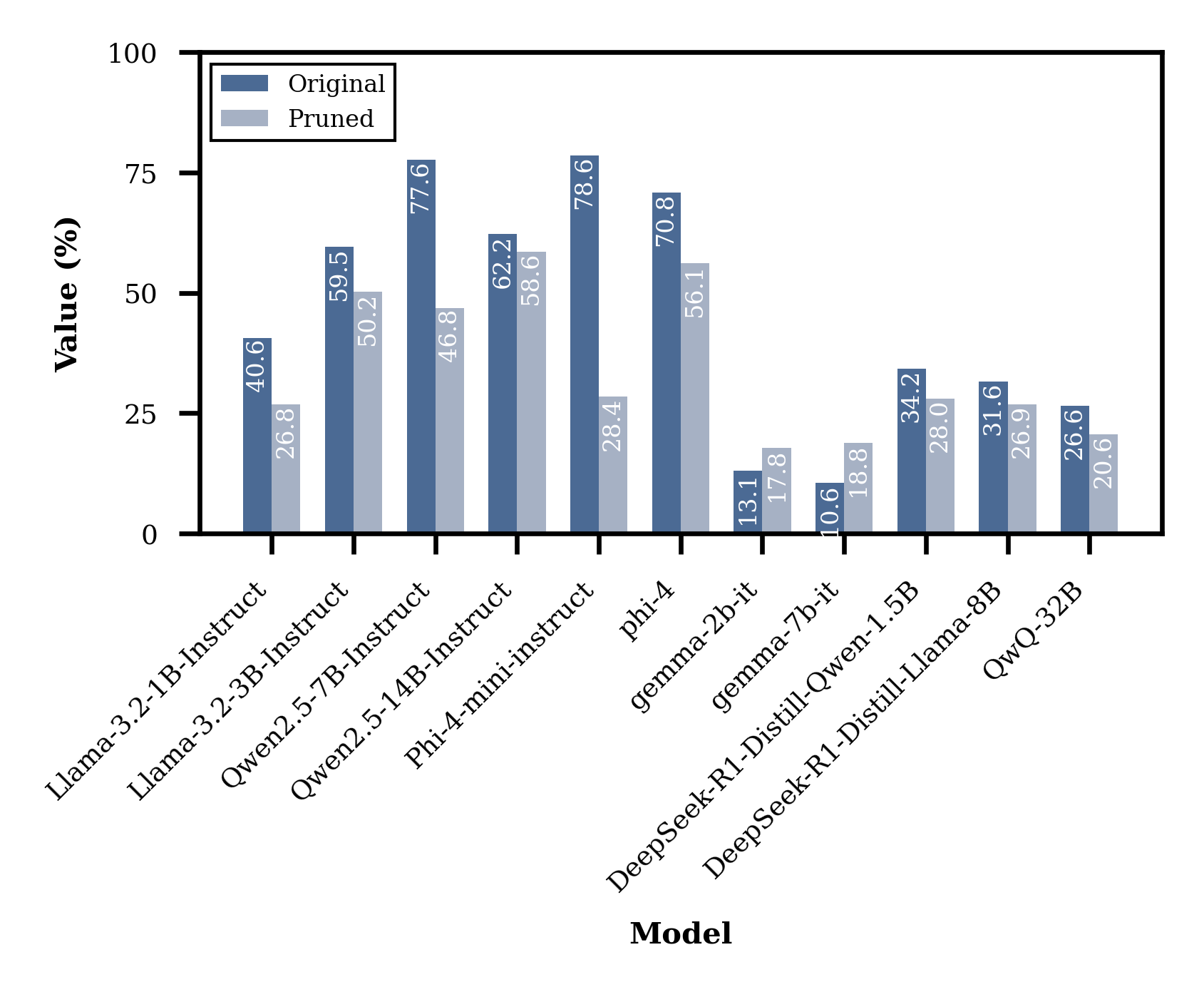}
    }
    \hfill
    \subfloat[CoLA]{%
        \includegraphics[width=0.32\textwidth]{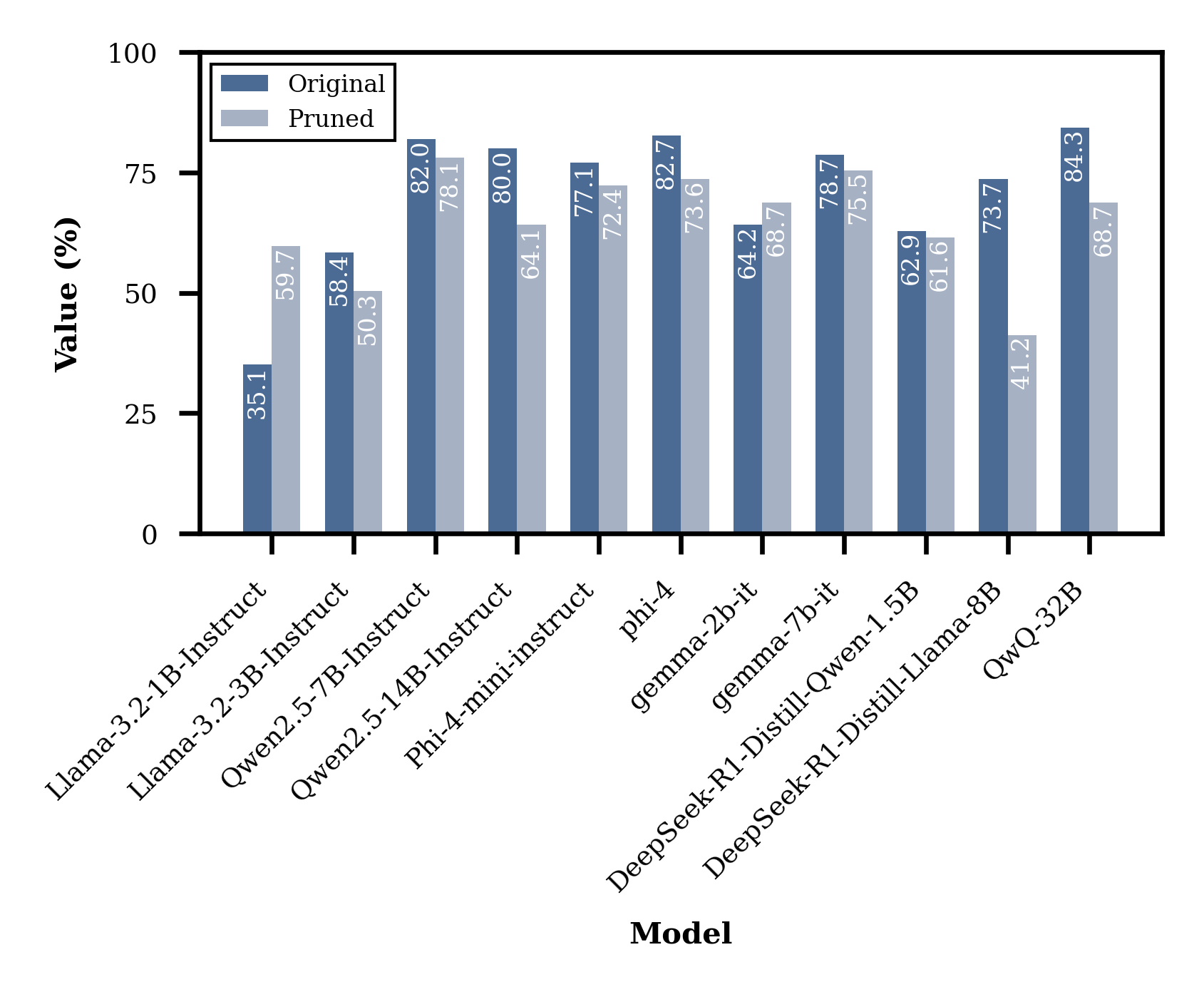}
    }

    \caption{Utility evaluation of original vs. pruned models across six NLU benchmarks with $z=2$.}
    \label{fig:utility-analysis-z2}
\end{figure*}

\begin{figure*}[htbp]
    \centering
    \subfloat[HellaSwag]{%
        \includegraphics[width=0.32\textwidth]{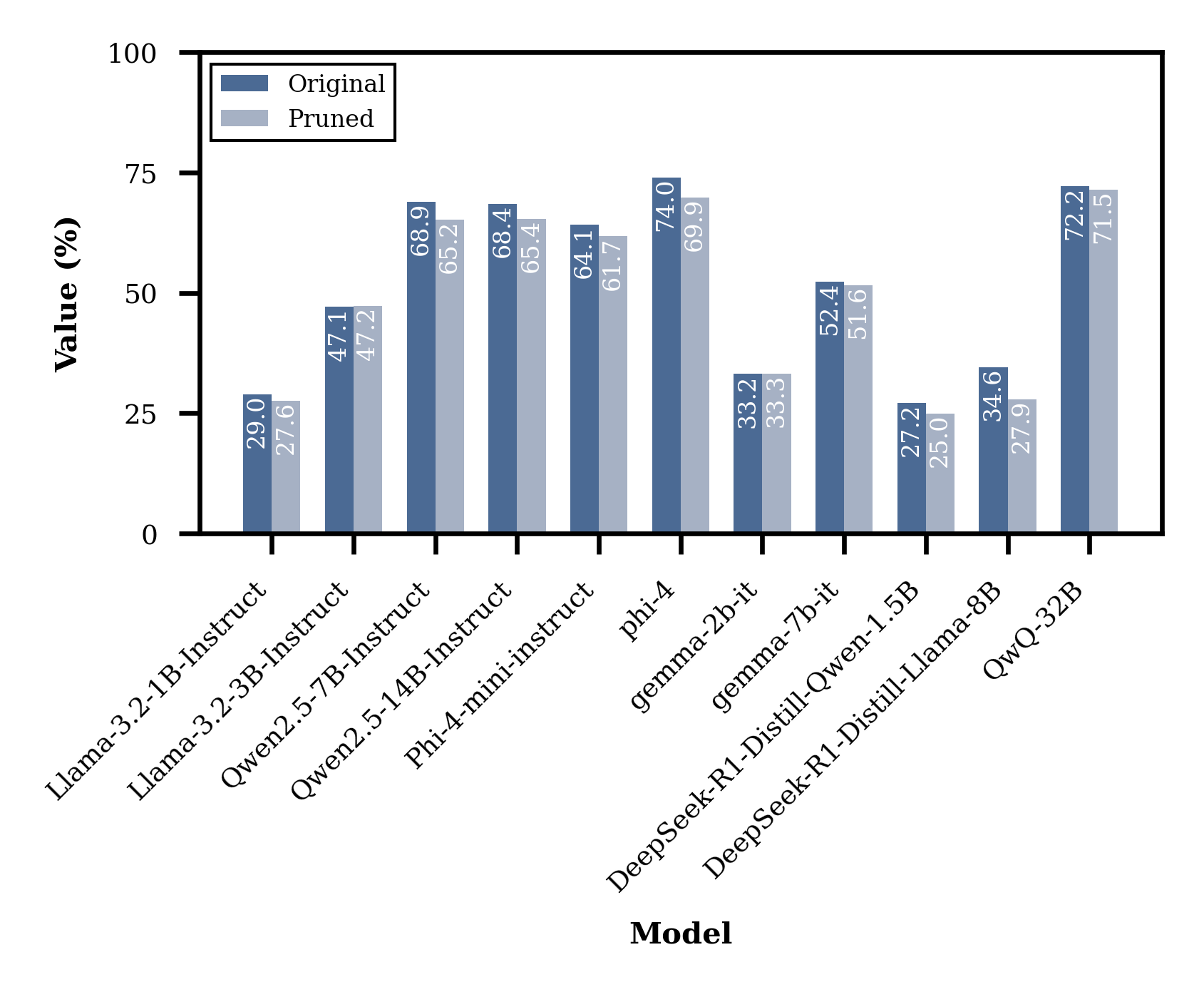}
    }
    \hfill
    \subfloat[RTE]{%
        \includegraphics[width=0.32\textwidth]{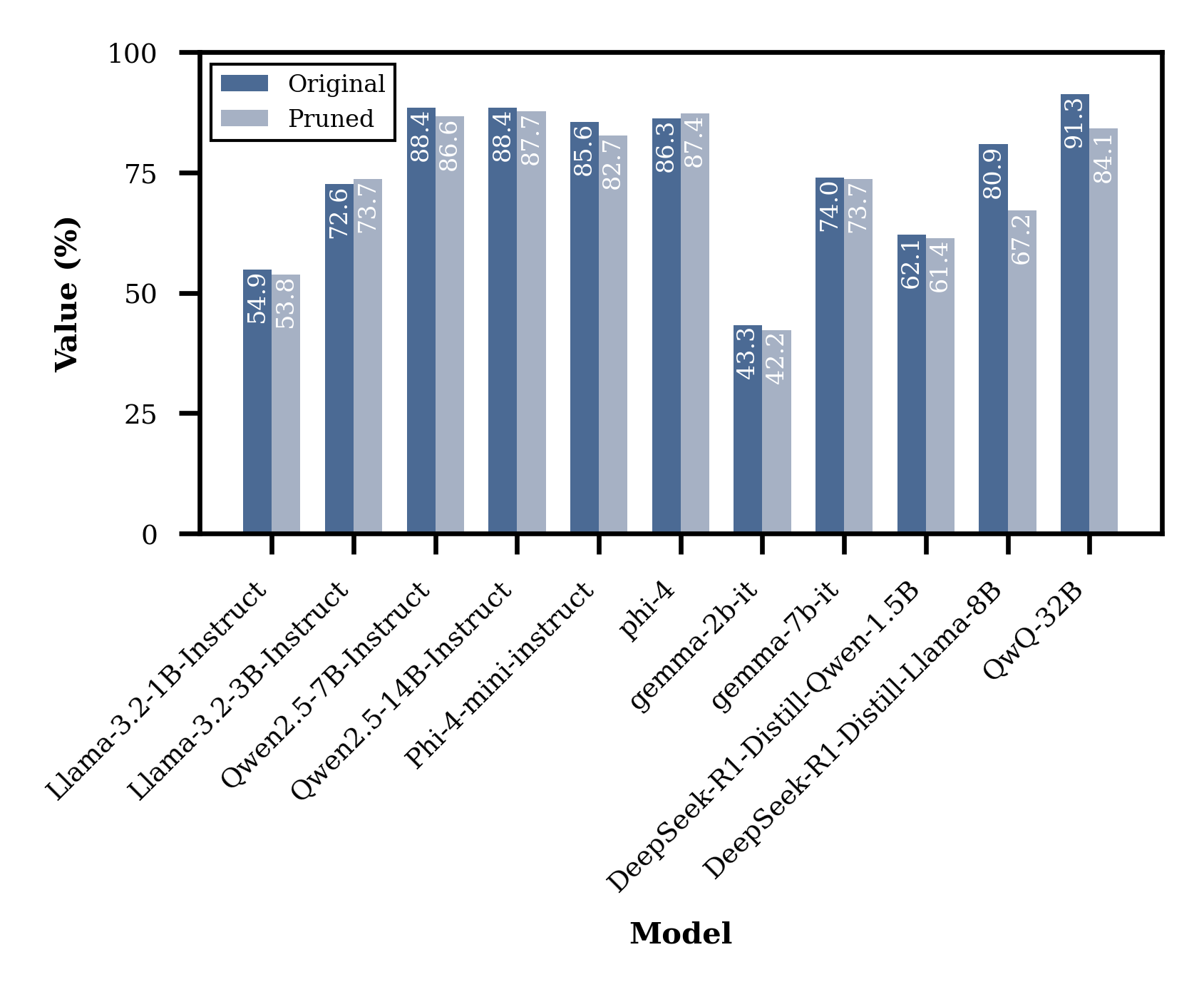}
    }
    \hfill
    \subfloat[WinoGrande]{%
        \includegraphics[width=0.32\textwidth]{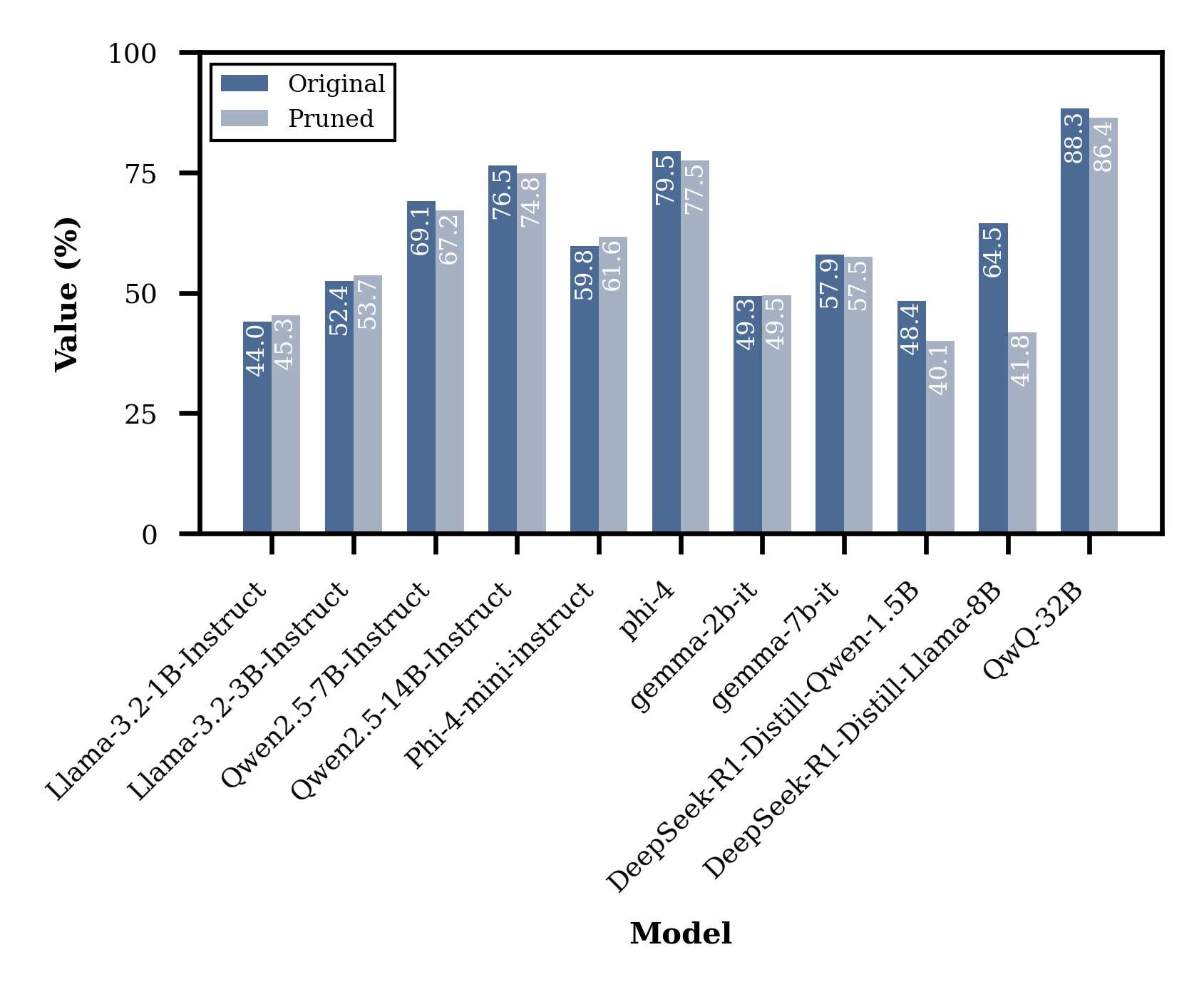}
    }

    \vspace{0.4em}

    \subfloat[ARC Challenge]{%
        \includegraphics[width=0.32\textwidth]{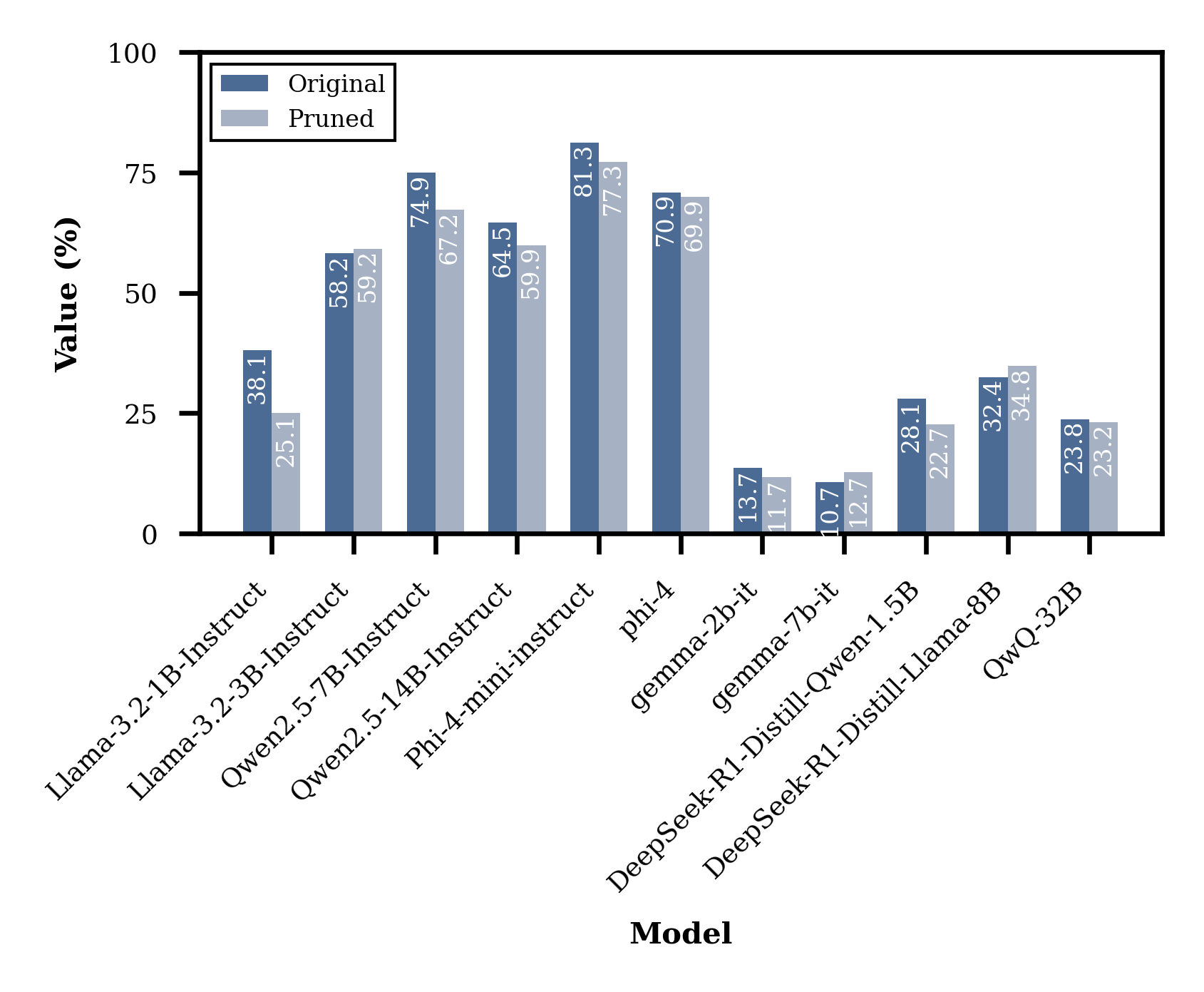}
    }
    \hfill
    \subfloat[OpenBookQA]{%
        \includegraphics[width=0.32\textwidth]{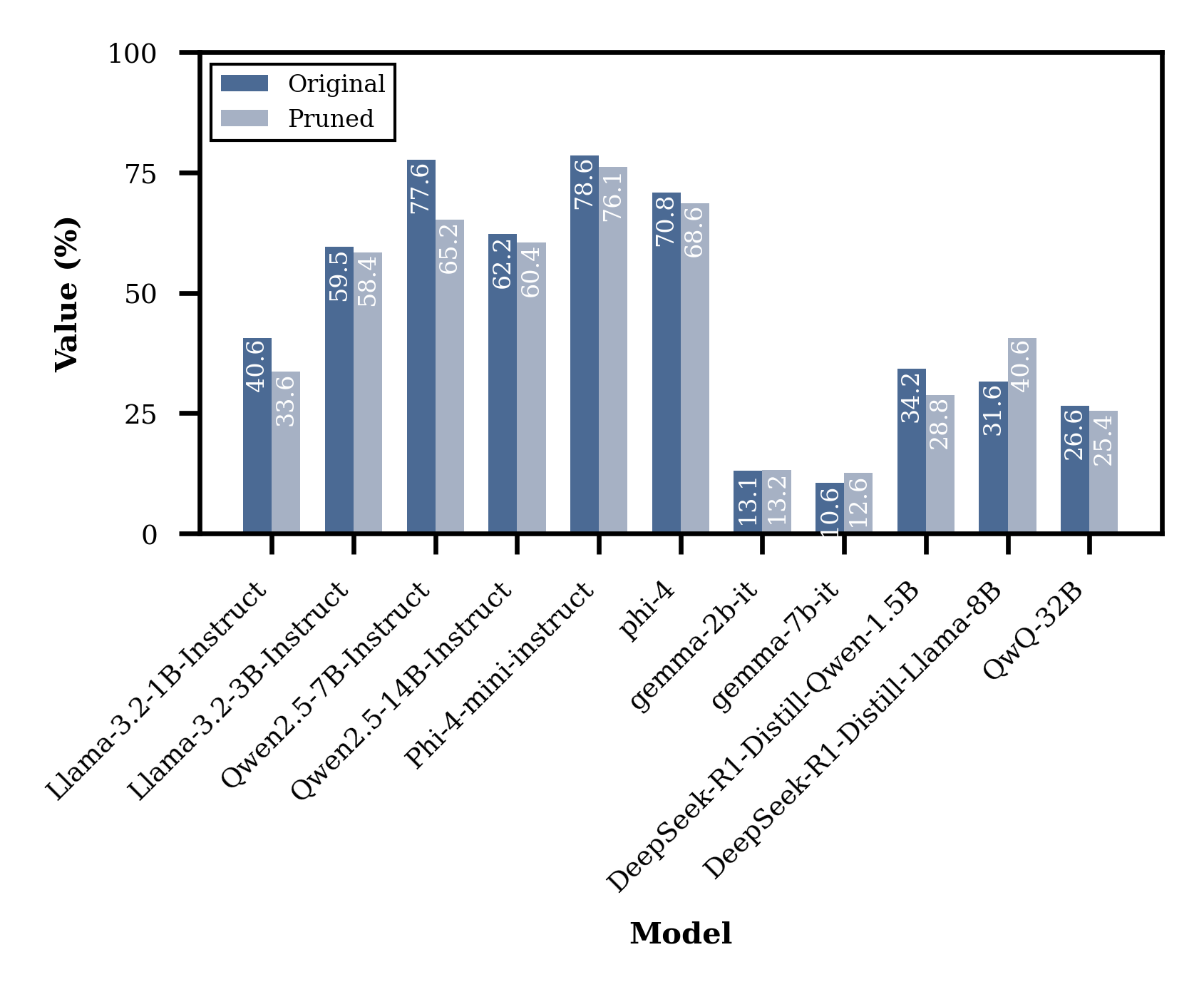}
    }
    \hfill
    \subfloat[CoLA]{%
        \includegraphics[width=0.32\textwidth]{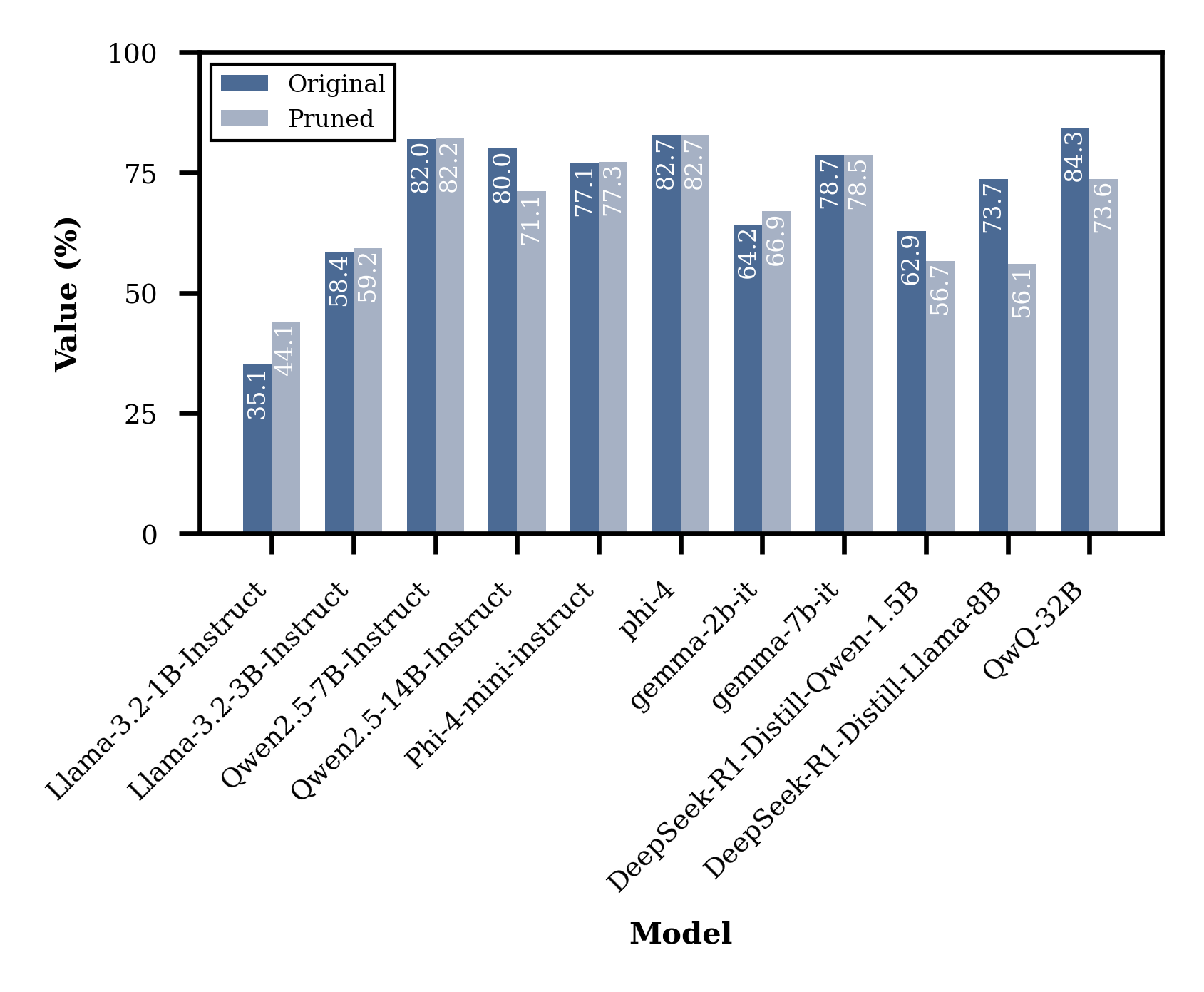}
    }

    \caption{Utility evaluation of original vs. pruned models across six NLU benchmarks with $z=4$.}
    \label{fig:utility-analysis-z4}
\end{figure*}

\newpage
\section{Artifacts}
\label{apdx:artifact}
\subsection{Description \& Requirements}

\ourname is a neuron-level attack framework designed to disable safety alignment in large language models (LLMs). This artifact fully supports the experiments and findings presented in the paper by supplying all necessary code and detailed instructions to replicate both white-box and black-box attack pipelines. The artifact provides scripts for identifying and pruning safety neurons in white-box models, generating jailbreak prompts through supervised fine-tuning, scoring neuron activations, and profiling LLMs to transfer safety vulnerabilities to proprietary black-box targets.

\subsection{How to access}
The complete artifact is hosted at the permanent archival repository: \url{https://doi.org/10.5281/zenodo.17072075}. The repository contains source code, documentation, and the environment specification file (\textit{environment.yml}) to reproduce all experimental results.

\subsection{Hardware dependencies}
White-box attacks can be executed on CPUs, but we strongly recommend using CUDA-enabled GPUs for practical runtimes. Black-box experiments, which involve fine-tuning and large-scale inference, require one or multiple GPUs. All evaluations in the paper were conducted using NVIDIA A100 and H100 GPUs; however, any modern GPU with at least 24 GB of VRAM should be sufficient for reproducing white-box attack results.

\subsection{Software dependencies}
The artifact is tested on Ubuntu 24.04 LTS with Python 3.10.16. Conda 24.11.3 is adopted for environment management. Dependencies include PyTorch (with CUDA), HuggingFace Transformers and Datasets, as well as auxiliary libraries such as \texttt{accelerate}, \texttt{bitsandbytes}, and \texttt{peft}. All packages are listed in \textit{environment.yml}, and can be installed in a single step using Conda.

\subsection{Benchmark}
The artifact evaluates over 30 open-weight LLMs from providers such as Meta (LLaMA), Google (Gemma), Microsoft (Phi), DeepSeek, and Alibaba (Qwen). Models must be downloaded via the HuggingFace model hub, and appropriate access must be requested where required. Due to ethical concerns, we do not release precomputed neuron activations or jailbreak prompt logs. Finetuned models for the black-box attack generator are also excluded due to size constraints, but can be reproduced using the provided training scripts.

\subsection{Artifact Installation \& Configuration}
To install the artifact, download and extract the repository, then navigate to the root directory. Ensure Conda is installed and execute:
\begin{lstlisting}[language=bash]
  $ conda env create -f environment.yml
  $ conda activate venv_neurostrike
\end{lstlisting}

This installs all required dependencies. GPU users must ensure the correct CUDA version is installed and that it is visible to PyTorch. All model and log paths are defined relative to the repository root.

\subsection{Experiment Workflow}
The artifact enables two primary workflows: (1) white-box attacks via pruning of identified safety neurons, and (2) black-box attacks via safety profiling and jailbreak prompt generation. The white-box workflow identifies safety neurons in an open-weight model, prunes them, and evaluates attack success rate (ASR) under varying thresholds. The black-box workflow trains a jailbreak prompt generator and neuron-level scorer, uses them to profile surrogate open-weight models, and transfers learned vulnerabilities to proprietary LLMs.

\subsection{Major Claims}
\begin{itemize}
    \item \textbf{C1}: \ourname disables safety mechanisms in white-box LLMs by pruning sparse and specialized safety neurons, achieving high ASR across multiple architectures, sizes, and families. This is supported by Experiment E1, with results shown in Table~\labelcref{tab:attacks tt llms}, Table~\labelcref{tab:attacks mm llms}, Table~\labelcref{tab:attacks ft llms}, and Table~\labelcref{tab:attacks dt llms}.
    \item \textbf{C2}: Our profiling method enables effective black-box jailbreak attacks via neuron-level knowledge transfer. This is validated by Experiment E2, with results presented in Section~\labelcref{sec:black-box evaluation} and Table~\labelcref{tab:asr_heatmap}.
\end{itemize}

\subsection{Evaluation}
\subsubsection{Experiment (E1): White-box Attacks}
\begin{itemize}
    \item Preparation: Activate the Conda environment and navigate to the \textit{white\_box} directory. Ensure that one has access to the desired model via HuggingFace.
    \item Execution: First, run \texttt{1\_get\_safety\_neuron.py} to identify safety neurons. Then execute \texttt{2\_prune\_and\_get\_asr.py} to prune the model and evaluate its ASR using adversarial prompts.
    \item Results: Logs will show the ASR under different pruning thresholds, matching Table~\labelcref{tab:attacks tt llms} and Table~\labelcref{tab:attacks mm llms} in the paper. The whole process is expected to take 10 human minutes and 100 compute minutes with a high-performance GPU. Runtime may vary depending on model size.
\end{itemize}

\subsubsection{Experiment (E2): Black-box Attacks}
\begin{itemize}
    \item Activate the environment and go to the \textit{black\_box} directory. Download \texttt{google/gemma-3b-it} via HuggingFace.
    \item Execution: Execute \texttt{1\_train\_generator.py} to train the jailbreak prompt generator. Then, run \texttt{2\_train\_scorer.py} to train the safety neuron scorer. Use \texttt{3\_profiling.py} to score neurons in surrogate models. Finally, attack the black-box LLM using \texttt{4\_attack.py}.
    \item Results: Generated prompts and logs are saved in \textit{\_black\_box\_jb\_data} and \textit{\_logs}. respectively. Evaluating these prompts on proprietary LLM APIs will reproduce Table~\labelcref{tab:asr_heatmap}. GPU runtime for training and inference across steps totals approximately 48 compute hours or more, depending on the experimental settings and computation resources.
\end{itemize}

\subsection{Customization}
To target different open-weight models, modify the \texttt{model\_id} field in the configuration files. Users can adjust the number of pruned neurons to control attack strength. In black-box workflows, prompt templates and scoring metrics can be adjusted and customized to suit different models or evaluation setups. The artifact is modular and supports seamless extension to new architectures or datasets.

\end{document}